\documentclass[manuscript=article]{achemso}
\pdfoutput=1
\usepackage{lineno}
\usepackage{subfigure}
\usepackage{amsmath}
\usepackage{float}
\usepackage{graphicx}
\usepackage{multirow}
\usepackage{longtable}
\usepackage[section]{placeins}
\usepackage{booktabs}

\modulolinenumbers[5]

\title[Anisotropic work function of elemental crystals]{Anisotropic work function of elemental crystals}
\author{Richard Tran}
\affiliation[UCSD]{Department of NanoEngineering, University of California San Diego, 9500 Gilman Dr, Mail Code 0448, La Jolla, CA 92093-0448, United States}
\author{Xiang-Guo Li}
\affiliation[UCSD]{Department of NanoEngineering, University of California San Diego, 9500 Gilman Dr, Mail Code 0448, La Jolla, CA 92093-0448, United States}
\author{Joseph Montoya}
\affiliation[LBL]{Environmental Energy Technologies Division, Lawrence Berkeley National Laboratory, Berkeley,  CA 94720, United States}
\author{Donald Winston}
\affiliation[LBL]{Environmental Energy Technologies Division, Lawrence Berkeley National Laboratory, Berkeley,  CA 94720, United States}
\author{Kristin Aslaug Persson}
\affiliation[LBL]{Environmental Energy Technologies Division, Lawrence Berkeley National Laboratory, Berkeley,  CA 94720, United States}
\alsoaffiliation[UCB]{Department of Materials Science \& Engineering, University of California Berkeley, Berkeley,  CA 94720-1760, United States}
\author{Shyue Ping Ong}
\email{ongsp@eng.ucsd.edu}
\affiliation[UCSD]{Department of NanoEngineering, University of California San Diego, 9500 Gilman Dr, Mail Code 0448, La Jolla, CA 92093-0448, United States}
\date{}

\begin{document}

\maketitle

\begin{abstract}
The work function is a fundamental electronic property of a solid that varies with the facets of a crystalline surface. It is a crucial parameter in spectroscopy as well as materials design, especially for technologies such as thermionic electron guns and Schottky barriers. In this work, we present the largest database of calculated work functions for elemental crystals to date. This database contains the anisotropic work functions of more than 100 polymorphs of about 72 elements and up to a maximum Miller index of two and three for non-cubic and cubic crystals, respectively. The database has been rigorously validated against previous experimental and computational data where available. We also propose a weighted work function based on the Wulff shape that can be compared to measurements from polycrystalline specimens, and show that this weighted work function can be modeled empirically using simple atomic parameters. Furthermore, for the first time, we were able to analyze  simple bond breaking rules for metallic systems beyond a maximum Miller index of one, allowing for a more generalized investigation of work function anisotropy.
\end{abstract}

\section{Introduction}

The work function ($\Phi$) is an electronic surface property of crystalline solids and is crucial to the understanding and design of materials in many applications. It can be directly applied to the engineering of device specifications such as the Schottky barrier of semiconductor junctions or the thermionic currents of electron guns. Furthermore, it has been used to guide the engineering of interfacial interactions between metals and monolayer structures for nanoscale self-assembly\cite{Heimel2006}. The work function is also an important parameter in characterization techniques where it can influence the tip tunneling current of scanning tunneling microscopes or correct the binding energy in photo-electron spectroscopy (PES).

The work function has also been explored as a parameter for materials design. For example, previous experimental and computational investigations of Ni-alloys by \citet{Lu2013, Lu2016} have established a correlation between the work function and various mechanical properties such as toughness, hardness, ductility and bulk modulus. A more recent study using first-principle calculations found similar correlations for elemental crystalline solids\cite{Hua2016}. The work function has also been proposed as a possible parameter for the desorption rate of surface adsorbates\cite{Kawano2008}. Calculated work functions of hcp materials have also been used to screen for more effective metallic photocathodes.\cite{Li2015a} 

Much effort has also been devoted to modelling $\Phi$ itself. \citet{Michaelson1978} and~\citet{Miedema1973}, for example, were previously successful in modelling the polycrystalline work function as a linear function of electronegativity. The modeling of the anisotropic work function ($\Phi_{\mbox{\scriptsize{hkl}}}$) as a function of surface morphology and chemical environment has also garnered much attention. Smoluchowski smoothing is one such model which describes the contributions to the work function of metals as a result of isotropic electron spreading and anisotropic electron smoothing\cite{Smoluchowski1941}. The spreading of negative charges increases the work function while the anisotropic smoothing of negative charges at the surface decreases the work function. Smoothing increases with surface roughness (defined here as the reciprocal of the surface packing fraction\cite{Sokolov1984}) which decreases the work function. This model is supported by previous observations that the anisotropic surface energy ($\gamma_{\mbox{\scriptsize{hkl}}}$) is inversely proportional to $\Phi_{\mbox{\scriptsize{hkl}}}$ via the broken bond surface density\cite{Wang2014c, Ji2016}. Similarly, the Brodie model attempts to explain $\Phi_{\mbox{\scriptsize{hkl}}}$ for transition metals as a function of (bulk) electron effective mass, surface atomic radius and inter-planar distance\cite{Brodie1995, Wojciechowski1997}. A more recent model using a dielectric formalism has been proposed by \citet{Fazylov2014} that describes $\Phi_{\mbox{\scriptsize{hkl}}}$ using surface roughness and surface plasmon dispersion.

An extensive database for $\Phi_{\mbox{\scriptsize{hkl}}}$ would be invaluable for validating and further expanding upon these models. However, experimentally measured work functions are usually for polycrystalline specimens ($\Phi_{\mbox{\scriptsize{poly}}}^{\mbox{\scriptsize{expt}}}$) instead of single crystals. An example of this is the extensive collection of experimentally measured $\Phi_{\mbox{\scriptsize{poly}}}^{\mbox{\scriptsize{expt}}}$ for 66 polycrystalline elemental solids compiled by \citet{Michaelson1977}. Though measurements for anisotropic $\Phi_{\mbox{\scriptsize{hkl}}}$ are not uncommon, values often vary due to the many techniques used or non-standardized methods of implementing the same technique (e.g., PES)\cite{Derry2015,Helander2010}. The sparsity of $\Phi_{\mbox{\scriptsize{hkl}}}$ and the lack of a comprehensive compilation with a single standardized technique makes it difficult to develop and gain insights into work function anisotropy using experimental measurements. 

Here, density functional theory (DFT) has the advantage of calculating $\Phi_{\mbox{\scriptsize{hkl}}}$ for a model of any specific solid surface under a controllable set of parameters, making it possible to create a standardized collection of values. Many authors have attempted such compilations for $\Phi_{\mbox{\scriptsize{hkl}}}$, which are often times accompanied by the corresponding surface energy $\gamma_{\mbox{\scriptsize{hkl}}}$\cite{Singh-Miller2009, Ji2016, Methfessel1992, Waele2016, Skriver1992, Wang2014c}. For instance, \citet{Ji2016} and \citet{Wang2014c} have calculated $\Phi_{\mbox{\scriptsize{hkl}}}$ for numerous bcc, fcc and hcp materials. \citet{Waele2016} created a database of $\Phi_{\mbox{\scriptsize{hkl}}}$ for all elemental crystalline solids, but only for facets up to a max Miller index (MMI) of 1, using the Perdew-Burke-Ernzerhof generalized gradient approximation (PBE-GGA) and localized density approximation (LDA) functionals. More recently, \citet{Patra2017} evaluated the performance of various functionals by calculating $\Phi_{\mbox{\scriptsize{hkl}}}$ for an MMI of one for Al, Cu, Ru, Rh, Pd, Ag, Pt and Au. Despite the wide variety of computational data, the majority of these studies are limited to small Miller indices (typically MMI of 1). In addition, computational data for lanthanide systems and different polymorphs is sparse\cite{Durakiewicz2001, Alden1995, Skriver1992}. Furthermore, most compilations do not consider possible reconstructions, which can drastically affect the calculated work function\cite{Waele2016}.

Here, we report the development of a comprehensive, validated database of work functions for elemental crystalline solids using DFT calculations that addresses all the above limitations in the following ways:
\begin{enumerate}
\item Coverage of 142 polymorphs of 72 elements, including rare earth metals.
\item Facets up to an MMI of three and two for cubic and non-cubic crystals, respectively, are considered.
\item Common reconstruction schemes, such as the missing-row (110) fcc and the diamond-type reconstructions,\cite{Stekolnikov2002b} have been taken into account.
\end{enumerate}
We validate our computed work functions with past experimental and computational data for both $\Phi_{\mbox{\scriptsize{hkl}}}$ and $\Phi_{\mbox{\scriptsize{poly}}}^{\mbox{\scriptsize{expt}}}$. Finally, we will discuss trends in the work function of the elements, and develop a predictive empirical model for $\Phi_{\mbox{\scriptsize{poly}}}^{\mbox{\scriptsize{expt}}}$.

\section{Methods}

\subsection{Definitions}

\begin{figure}[H]
\subfigure{\includegraphics[width=0.9\textwidth]{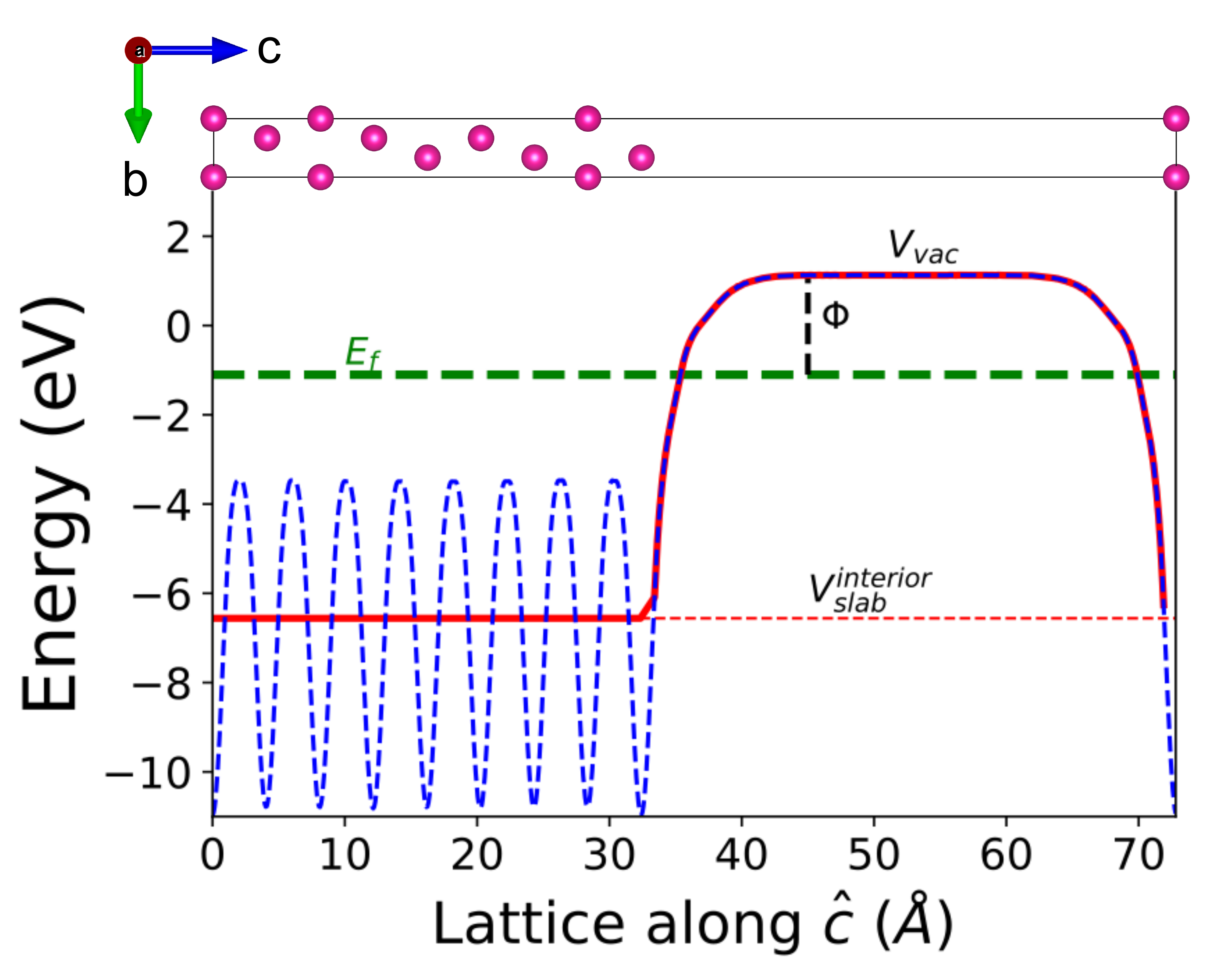}}
\caption{\label{fig:potential_v_energy_w_lattice} Plot of the electrostatic potential along the hcp Rb (0001) slab model. The Fermi energy ($E_{\mbox{\scriptsize{f}}}$),  electrostatic potential of the vacuum region ($V_{\mbox{\scriptsize{vac}}}$), average electrostatic potential of the slab region ($V^{\mbox{\scriptsize{interior}}}_{\mbox{\scriptsize{slab}}}$) and work function ($\Phi$) are indicated.}
\end{figure}

The work function is defined as the energy barrier required to move an electron from the surface of a solid material into free space, as given by the following expression:
\begin{equation}
\label{eqn:work_function_eqn}
\Phi=V_{\mbox{\scriptsize{vac}}} - E_{\mbox{\scriptsize{F}}}
\end{equation}
where $V_{\mbox{\scriptsize{vac}}}$ is the electrostatic potential of the vacuum region near the surface and $E_{\mbox{\scriptsize{f}}}$ is the Fermi energy of the slab. The energy barrier can be visualized in Figure~\ref{fig:potential_v_energy_w_lattice} where $V_{\mbox{\scriptsize{vac}}}$ is obtained when the electron is far enough away from the surface, that the potential remains constant over a small distance in the vacuum. This method has been widely used in previous studies for calculating the work function~\cite{Waele2016, Wang2014c, Ji2016, Skriver1992} and has been shown to converge quickly with respect to slab thickness\cite{Singh-Miller2009}.

\subsection{Modeling non-uniform work functions}

For comparison to work functions obtained from polycrystalline specimens, one approach is to calculate the work function of a ``patchy'' surface by weighting each $\Phi_{\mbox{\scriptsize{hkl}}}$ by the area fraction of its corresponding facet\cite{Woodruff2016, Kawano2008} as follows: 
\begin{equation}
\label{eqn:average_work function}
\bar{\Phi} = \frac{\sum_{\{hkl\}}\Phi_{\mbox{\scriptsize{hkl}}}A_{\mbox{\scriptsize{hkl}}}}{\Sigma A_{\mbox{\scriptsize{hkl}}}}=\sum_{\{hkl\}}\Phi_{\mbox{\scriptsize{hkl}}}f^A_{\mbox{\scriptsize{hkl}}}
\end{equation}
where $A_{\mbox{\scriptsize{hkl}}}$ and $f^{\mbox{\scriptsize{A}}}_{\mbox{\scriptsize{hkl}}}$ are the total area and the area fraction of all facets in the $\{hkl\}$ family, respectively. A polycrystal is an extreme case of a patchy surface, and as such the same technique can be applied to $\Phi_{\mbox{\scriptsize{poly}}}^{\mbox{\scriptsize{expt}}}$. The PES signal of the patch with a lower work function will tend to eclipse those with higher work functions leading to an underestimated measurement of $\Phi_{\mbox{\scriptsize{poly}}}^{\mbox{\scriptsize{expt}}}$. Thus, experimental values of the lowest anisotropic work function ($\Phi^{\mbox{\scriptsize{lowest}}}_{\mbox{\scriptsize{hkl}}}$) are only $\sim90$ meV lower than $\Phi_{\mbox{\scriptsize{poly}}}^{\mbox{\scriptsize{expt}}}$. Because of this, it has also been suggested that $\Phi^{\mbox{\scriptsize{lowest}}}_{\mbox{\scriptsize{hkl}}}$ is a good estimate of $\Phi_{\mbox{\scriptsize{poly}}}^{\mbox{\scriptsize{expt}}}$\cite{Helander2010, Waele2016, Kawano2008}. In this study, we use the facets present in the Wulff shape previously calculated by the current authors\cite{Tran2016a} as an estimate of the orientation and area fraction present in a polycrystalline sample to obtain $\bar{\Phi}$ and $\Phi^{\mbox{\scriptsize{lowest}}}_{\mbox{\scriptsize{hkl}}}$ as estimates for $\Phi_{\mbox{\scriptsize{poly}}}^{\mbox{\scriptsize{expt}}}$.

As mentioned earlier, Smoluchowski smoothing describes the anisotropic work function of metals as being inversely correlated with the broken bonds per surface area. As such, we model our values for $\Phi_{\mbox{\scriptsize{hkl}}}$ normalized by $\bar{\Phi}$ using the ratio of broken bonds-to-bulk coordination number ($\frac{N_{\mbox{\scriptsize{BB}}}}{CN_{\mbox{\scriptsize{bulk}}}}$) in a slab normalized by the surface area-to-atomic area ratio ($\frac{A_{\mbox{\scriptsize{surf}}}}{\pi r^{2}_{\mbox{\scriptsize{A}}}}$) in order to compare across all systems:
\begin{equation}
\label{eqn:normalized_broken_bonds}
\overline{\mbox{BB}} = \frac{N_{\mbox{\scriptsize{BB}}}}{CN_{\mbox{\scriptsize{bulk}}}}\times\frac{A_{\mbox{\scriptsize{surf}}}}{\pi r^{2}_{\mbox{\scriptsize{A}}}}
\end{equation}
It is known that for bcc and hcp materials, the first nearest neighbors (1NN) and second nearest neighbors (2NN) are in close proximity to each other, leading to contributions from the latter to the anisotropy of surface energy\cite{Frank1958}. Hence, when defining $\overline{\mbox{BB}}$ for a material, we will limit the maximum coordination number to the 1NN for fcc materials and explore both 1NN and 2NN for hcp and bcc materials. For hcp structures, we omitted $\Phi_{0001}$ when investigating the effect of the 1NN as the number of broken bonds will always be 0 which is unphysical. The inverse correlation between $\overline{\mbox{BB}}$ and $\Phi_{\mbox{\scriptsize{hkl}}}$ for each element can be quantified by the Pearson correlation coefficient ($r$). We will define systems with a negative linear trend between $\Phi_{\mbox{\scriptsize{hkl}}}$ and the normalized broken bonds with $r<-0.75$ as having a strong correlation, $-0.75<r<-0.5$ as having a moderate correlation and $r>-0.5$ as having a weak correlation. Only ground state metallic fcc, bcc and hcp systems were explored under this context.

\subsection{Computational details and workflow}
For all slab calculations, we performed a full relaxation of the site positions under a fixed volume before obtaining the electrostatic potential of the slabs (see ref \citenum{Tran2016a} for a complete description of computational details). The electrostatic potentials only contains the electrostatic contributions (no contributions from the exchange correlation). All calculations were performed using the Vienna Ab initio Simulation Package (VASP) with the exchange-correlation effects modeled using the PBE-GGA functional. Calculations using the revised PBE (rPBE) functional were performed on a smaller set of data using the same parameters for comparison.

We used the high-throughput workflow proposed by~\citet{Sun2013a} and implemented by~\citet{Tran2016a} and~\citet{Montoya2017} to obtain all required data. The workflow was implemented using the open-source software packages Python Materials Genomics,\cite{Ong2013} FireWorks\cite{Jain2015b} and Atomate\cite{Mathew2017}. The work function is extracted from the calculations and inserted into the same database. To handle errors that may arise during calculations, the custodian software package was used as a wrapper around VASP together with a set of robust error handling rules. The database will be continuously improved and will continue growing as more structural data becomes available on the Materials Project\cite{Winston2016}. 

\subsection{Data Availability}

The data can be accessed from the elemental-surface-data-focused Crystalium\cite{CrystaliumWebsite} website, as well as from the Materials Project website\cite{MPWebsite} on its detail pages for specific crystals.

\section{Results}

\begin{figure}[H]
\includegraphics[width=0.9\textwidth]{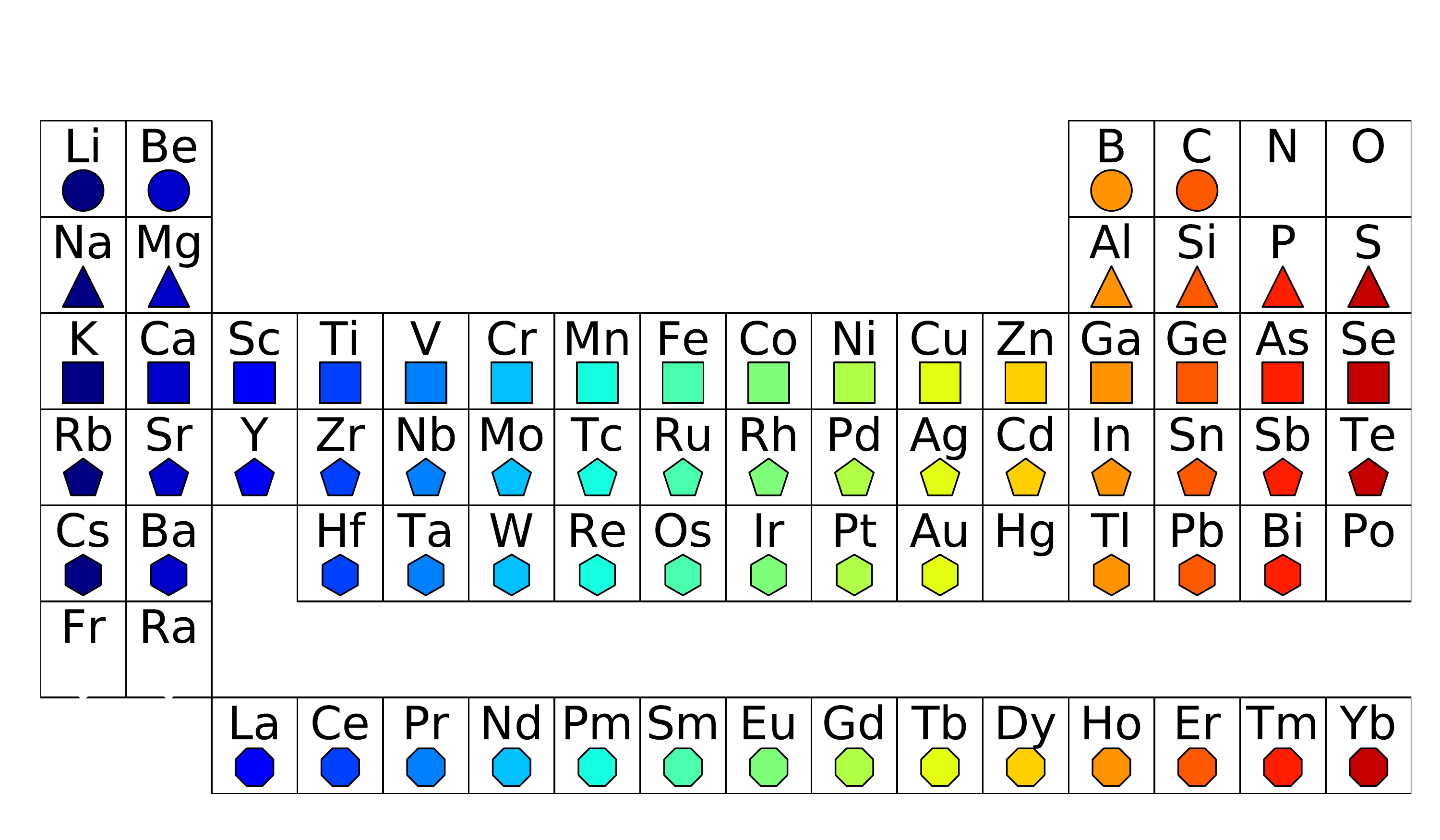}
\caption{\label{fig:ptlegend} Marker shape and color scheme for plots.}
\end{figure}

Due to the vast number of data points for $\Phi_{\mbox{\scriptsize{hkl}}}$ when comparing to literature values, we have adopted a consistent marker shape and color scheme for ease of reference (see Figure \ref{fig:ptlegend}) for all subsequent plots. The shape and color represents the row and group of the element in the periodic table, respectively.

All values for $\Phi$ reported in this study, including those found in the literature, are listed in the Supplementary Information in Tables~\ref{fig:PolyAverageTable} and~\ref{fig:XCComparisonTable}. Literature values for $\Phi$ were taken from the most recent experimental and computational studies available during the writing of this manuscript. Experimental values are explicitly annotated with a ``expt'' superscript, e.g., $\Phi_{\mbox{\scriptsize{poly}}}^{\mbox{\scriptsize{expt}}}$, and unless otherwise indicated, all other values are computed values.

\subsection{Experimental and computational validation}

\begin{figure}[H]
\subfigure[$\Phi_{\mbox{\scriptsize{poly}}}^{\mbox{\scriptsize{expt}}}$ vs $\bar{\Phi}$]
{\label{fig:dft_wulff_vs_exp_poly}\includegraphics[width=0.49\textwidth]{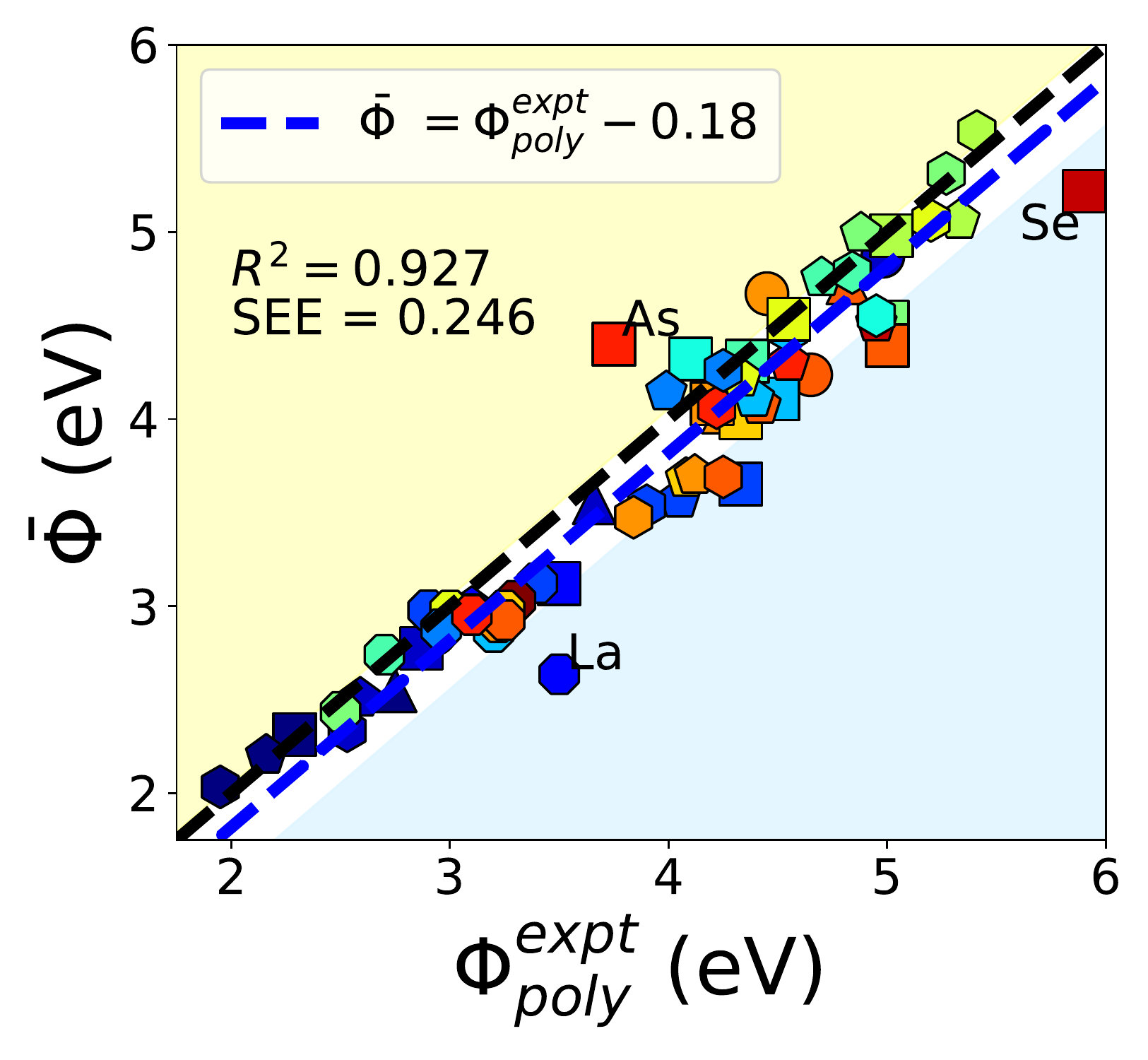}}
\subfigure[$\Phi_{\mbox{\scriptsize{poly}}}^{\mbox{\scriptsize{expt}}}$vs $\Phi^{\mbox{\scriptsize{lowest}}}_{\mbox{\scriptsize{hkl}}}$]
{\label{fig:dft_lowest_vs_exp_poly}\includegraphics[width=0.49\textwidth]{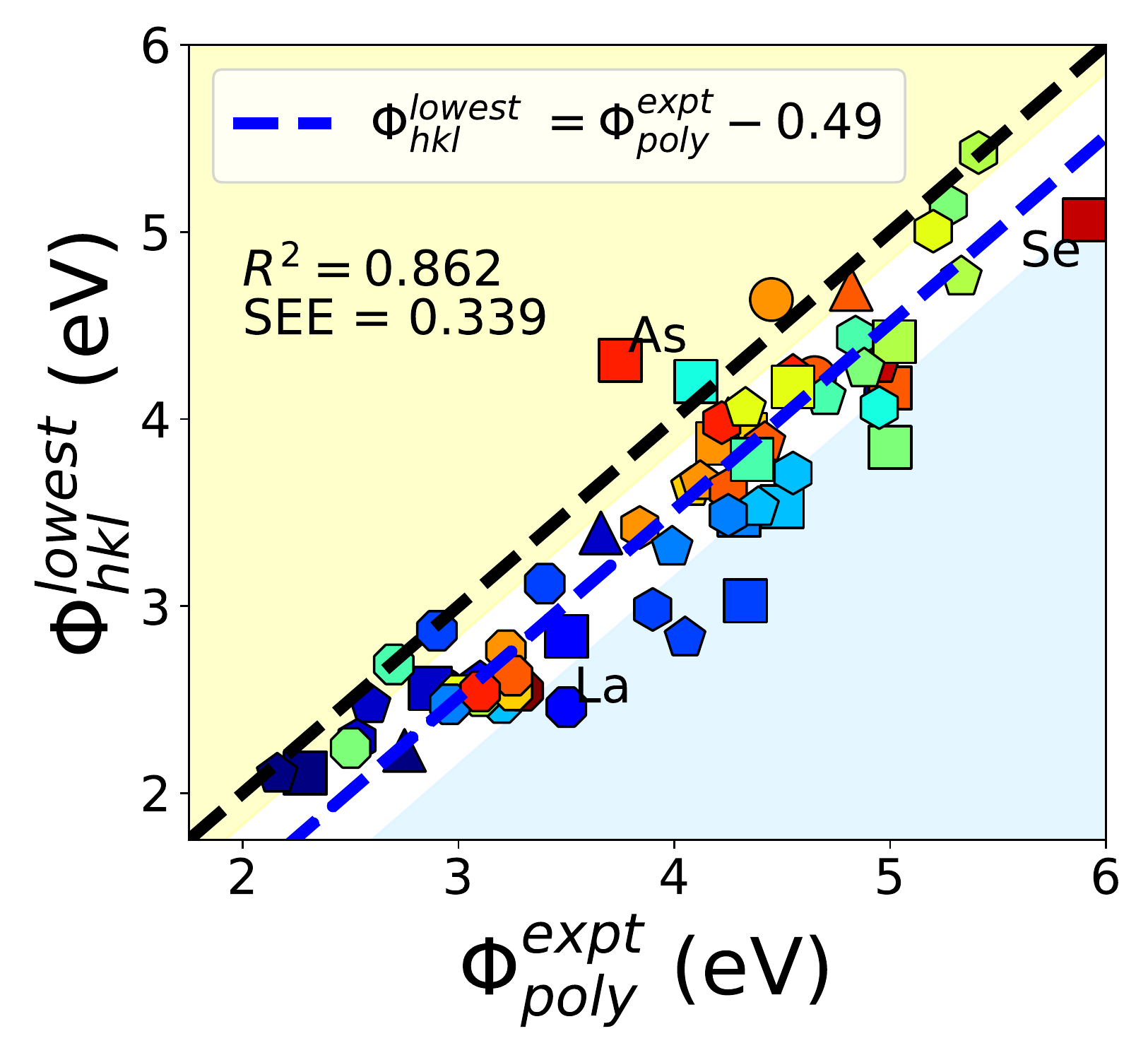}}
\caption{\label{fig:dft_wulff_lowest_vs_exp_poly} Plot of (a) experimentally measured $\Phi_{\mbox{\scriptsize{poly}}}^{\mbox{\scriptsize{expt}}}$ vs the computed $\bar{\Phi}$ and (b) $\Phi_{\mbox{\scriptsize{poly}}}^{\mbox{\scriptsize{expt}}}$ vs the computed $\Phi^{\mbox{\scriptsize{lowest}}}_{\mbox{\scriptsize{hkl}}}$. The single-factor linear regression line $y=x+c$ for both plots are indicated as dashed blue lines along with the $R^2$ value and standard error of the estimate (SEE). Values within the light blue (light yellow region) region are below (above) the SEE. (see refs \citenum{CRC, Kawano2008, Rozkhov1971}).}
\end{figure}

Figures~\ref{fig:dft_wulff_lowest_vs_exp_poly} shows a single-parameter $y = x + c$ least squares fit for $\bar{\Phi}$ vs $\Phi_{\mbox{\scriptsize{poly}}}^{\mbox{\scriptsize{expt}}}$ and $\Phi_{\mbox{\scriptsize{hkl}}}^{\mbox{\scriptsize{lowest}}}$ vs $\Phi_{\mbox{\scriptsize{poly}}}^{\mbox{\scriptsize{expt}}}$ for the ground state polymorph of each element. We find that the PBE $\bar{\Phi}$ are on average 0.31 eV closer to $\Phi_{\mbox{\scriptsize{poly}}}^{\mbox{\scriptsize{expt}}}$ than the PBE $\Phi_{\mbox{\scriptsize{hkl}}}^{\mbox{\scriptsize{lowest}}}$. The linear fit for $\bar{\Phi}$ vs $\Phi_{\mbox{\scriptsize{poly}}}^{\mbox{\scriptsize{expt}}}$ also yielded a higher $R^2$ of 0.927 and a lower standard error of the estimate (SEE) of 0.246 eV compared to that of $\Phi_{\mbox{\scriptsize{hkl}}}^{\mbox{\scriptsize{lowest}}}$ vs $\Phi_{\mbox{\scriptsize{poly}}}^{\mbox{\scriptsize{expt}}}$ ($R^2 = 0.862$ eV and SEE $= 0.339$ eV). We find that $\bar{\Phi}$   systematically underestimates $\Phi_{\mbox{\scriptsize{poly}}}^{\mbox{\scriptsize{expt}}}$ by 0.18 eV on average. Notable outliers include $\Phi^{\mbox{\scriptsize{expt,As}}}_{\mbox{\scriptsize{poly}}}$ which is underestimated, and $\Phi^{\mbox{\scriptsize{expt,La}}}_{\mbox{\scriptsize{poly}}}$ and $\Phi^{\mbox{\scriptsize{expt,Se}}}_{\mbox{\scriptsize{poly}}}$ which are overestimated by more than the SEE. The largest error is for La, with an error of 0.86 eV. Although the~\citet{Michaelson1977} values of $\Phi^{\mbox{\scriptsize{expt,Re}}}_{\mbox{\scriptsize{poly}}}$ is 0.44 eV closer to our calculated value, the more recent value reported by~\citet{Kawano2008} is reported here.

\begin{figure}[H]
\subfigure[$\Phi_{\mbox{\scriptsize{hkl}}}$ vs $\Phi^{\mbox{\scriptsize{expt}}}_{\mbox{\scriptsize{hkl}}}$ (literature).]
{\label{fig:all_wf_facets_vs_exp_facets}\includegraphics[width=0.49\textwidth]{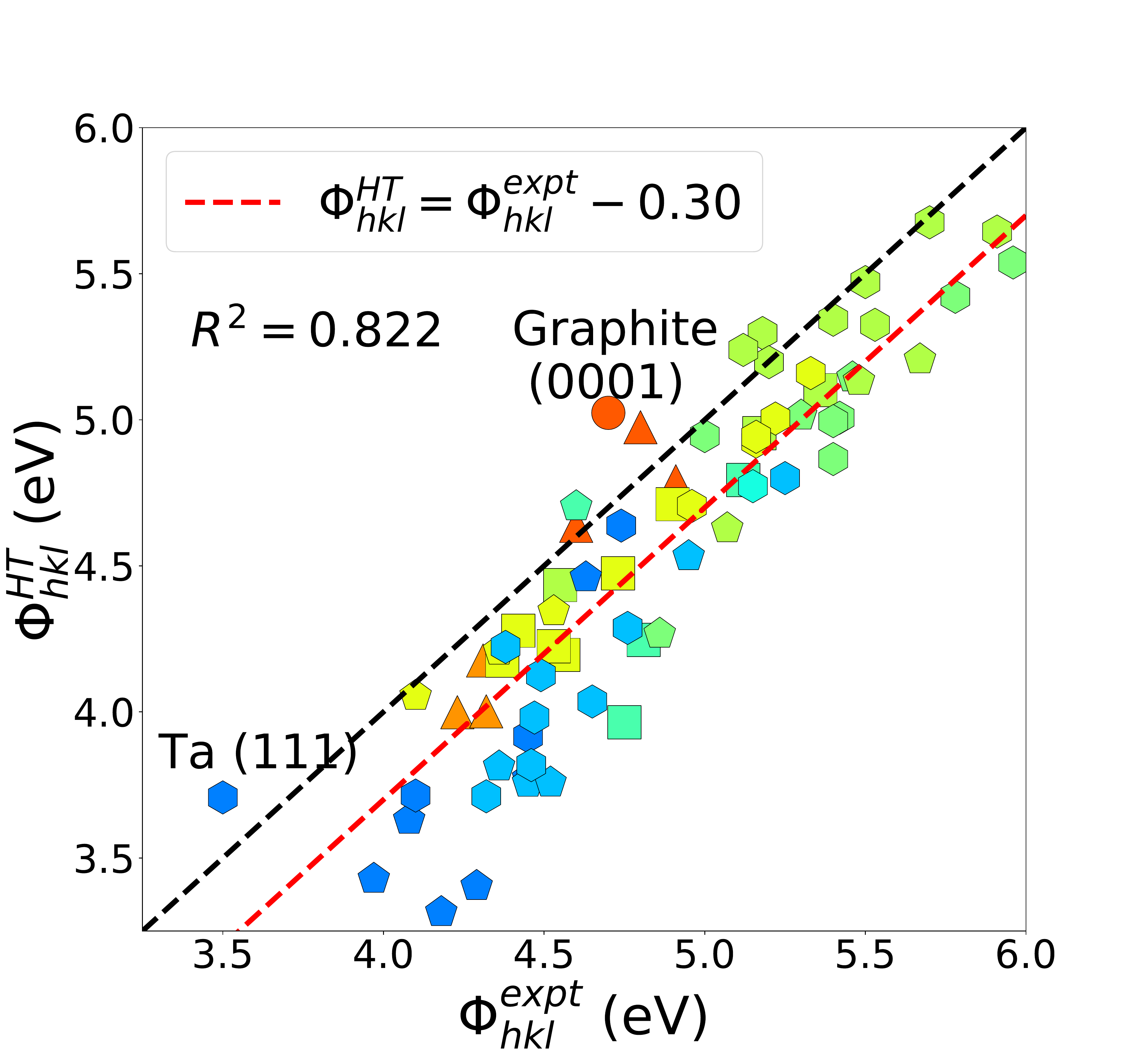}}
\subfigure[$\Phi_{\mbox{\scriptsize{hkl}}}$ vs PBE (literature).]
{\label{fig:all_wf_facets_vs_gga_facets}\includegraphics[width=0.49\textwidth]{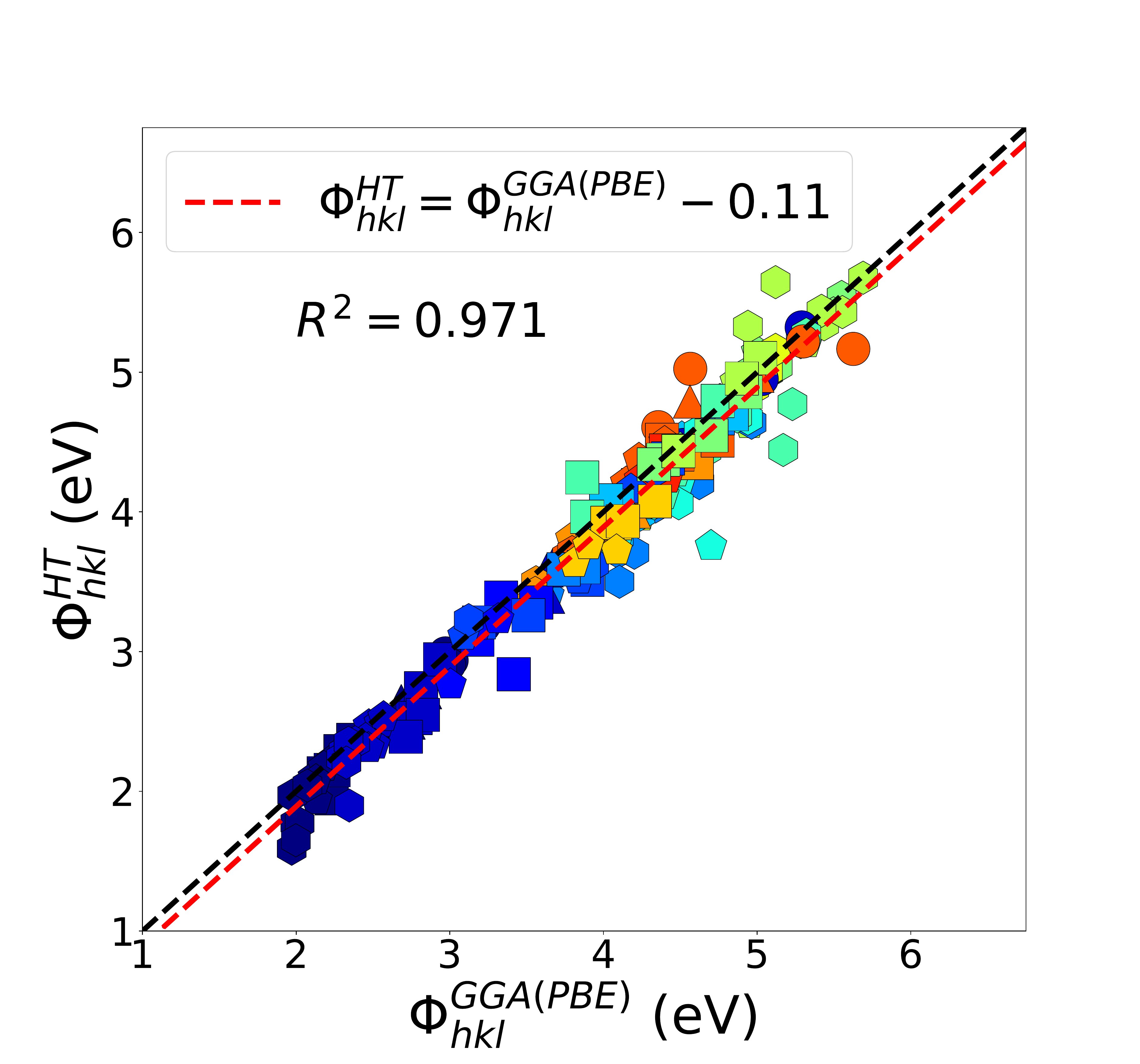}}
\quad
\subfigure[$\Phi_{\mbox{\scriptsize{hkl}}}$ vs LDA (literature).]
{\label{fig:all_wf_facets_vs_lda_facets}\includegraphics[width=0.49\textwidth]{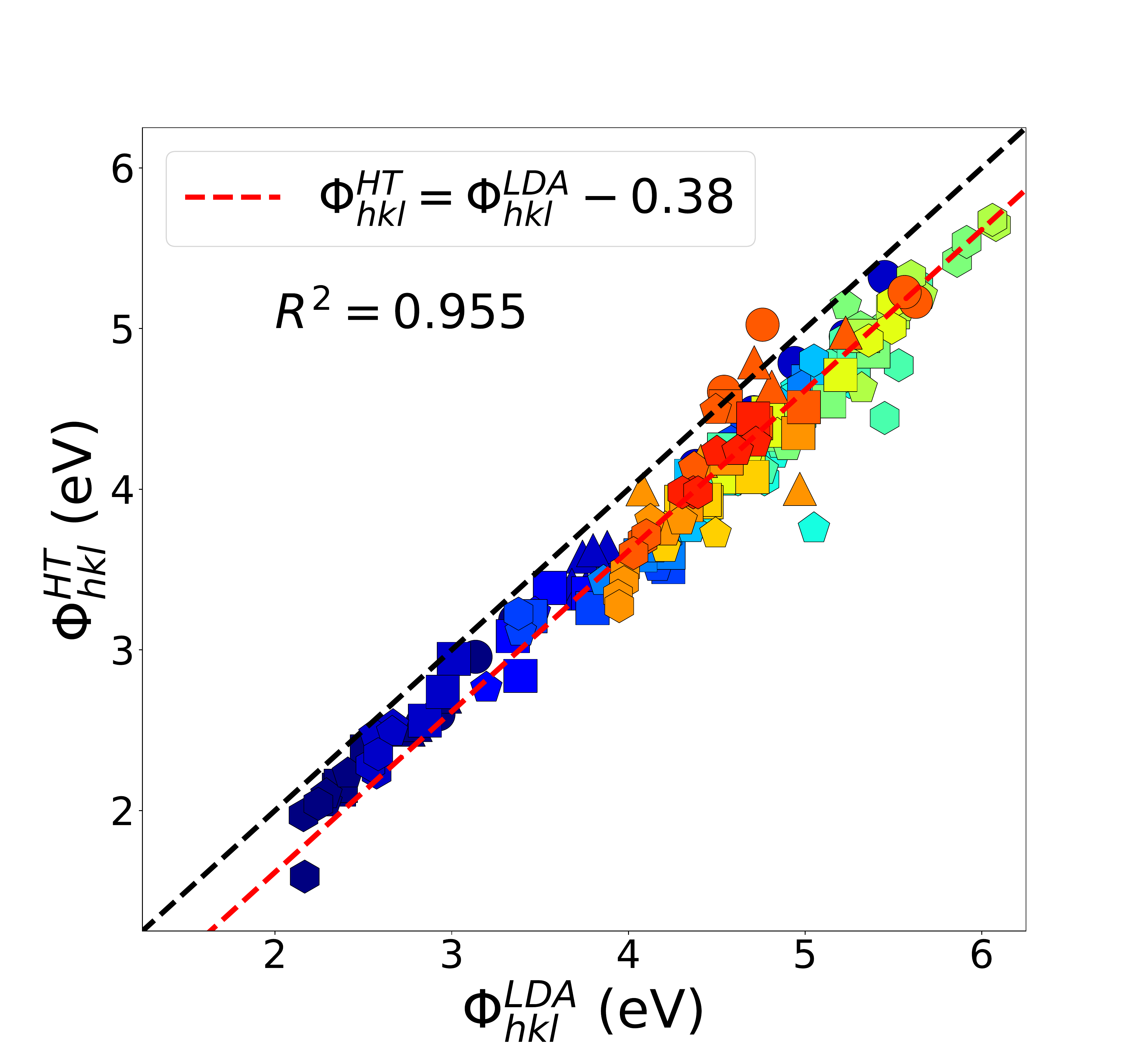}}
\subfigure[$\Phi_{\mbox{\scriptsize{hkl}}}$ vs RPBE (this study).]
{\label{fig:all_wf_facets_vs_rpbe_facets}\includegraphics[width=0.49\textwidth]{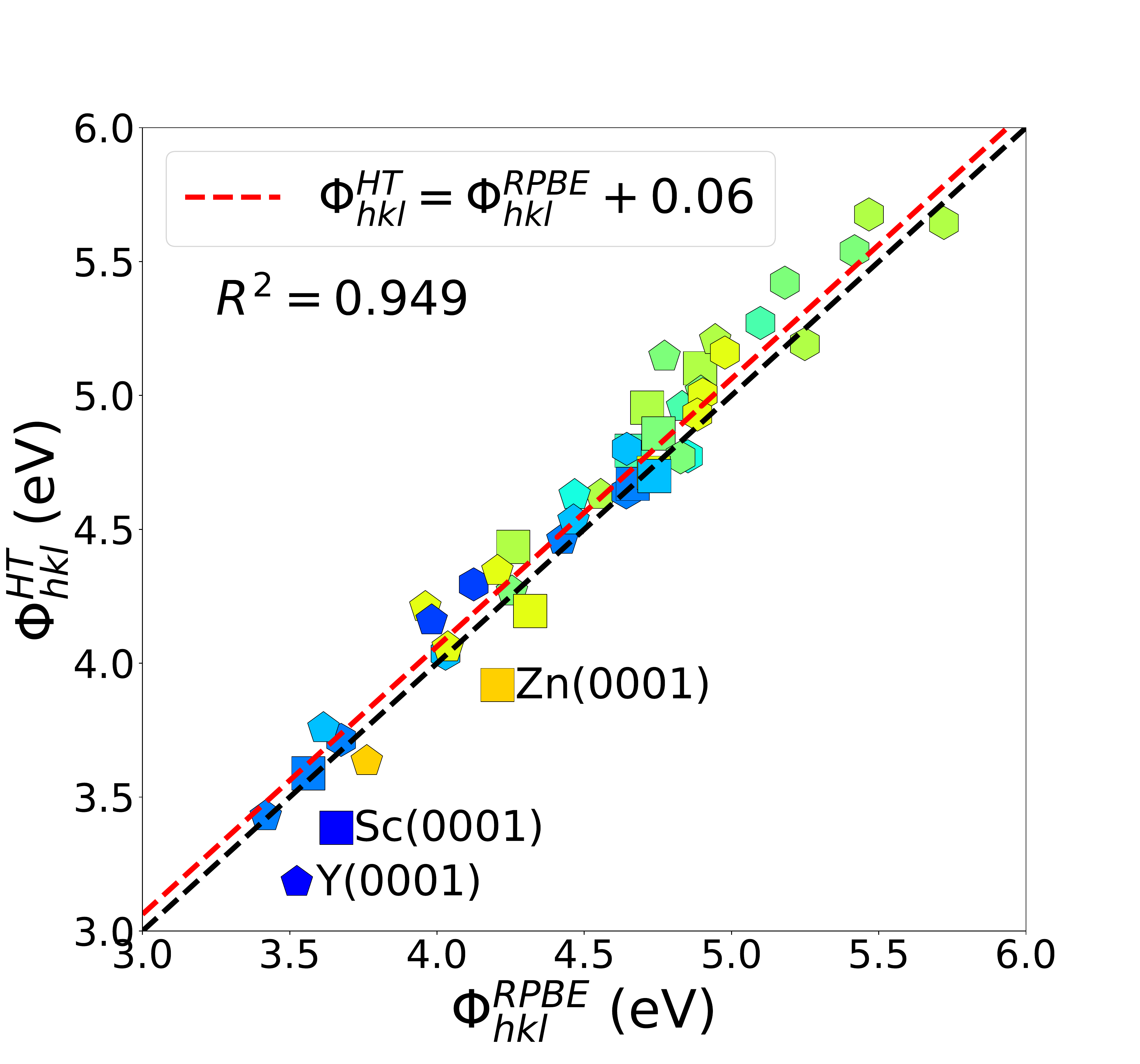}}
\caption{\label{fig:all_wf_facets_vs_xc} Plot of computed facet-dependent $\Phi^{\mbox{\scriptsize{HT}}}_{\mbox{\scriptsize{hkl}}}$ in this work vs (a) experimental values \cite{CRC, Kawano2008, Derry2015} (b) literature PBE values\cite{Waele2016, Ji2016, Wang2014c, Patra2017}, (c) literature LDA values\cite{Waele2016, Ji2016, Patra2017} and (d) RPBE values (this work).}
\end{figure} 

Figure~\ref{fig:all_wf_facets_vs_xc} compares the facet-dependent work functions obtained in this study to values obtained experimentally from single crystals and from other functionals, including LDA, PBE and RPBE. Again, we find that the experimental values in Figure~\ref{fig:all_wf_facets_vs_exp_facets} are on average 0.30 eV higher than the computed ones with an increasing deviation for work functions of lower values. The (100), (310) and (311) facets of the early transition and refractory metals (Mo, W and Nb) have some of the lowest work functions, which also have the greatest deviation between the PBE and experimental values. For the same elements, facets with higher work functions have a smaller deviation ((110) and (210)). 
An exception to this is $\Phi^{\mbox{\scriptsize{Ta}}}_{111}$ where computed and experimental values are in relative agreement despite having a value lower than other work functions. Meanwhile, the computed value of $\Phi^{\mbox{\scriptsize{Graphite}}}_{0001}$ greatly overestimates the experimental value (see ref \citenum{Kawano2008}). In general, the qualitative trends in work functions for different facets of each element are in agreement with the experimental trends, with the notable exception of Al. \citet{Eastmen1973} previously reported the order of $\Phi^{\mbox{\scriptsize{Al}}}_{\mbox{\scriptsize{hkl}}}$ to be $\Phi^{\mbox{\scriptsize{Al}}}_{111}>\Phi^{\mbox{\scriptsize{Al}}}_{100}>\Phi^{\mbox{\scriptsize{Al}}}_{110}$ which is typical of fcc metals while a later work by ~\citet{Grepstad1976} reported that $\Phi^{\mbox{\scriptsize{Al}}}_{100}>\Phi^{\mbox{\scriptsize{Al}}}_{110}\sim\Phi^{\mbox{\scriptsize{Al}}}_{111}$, which is consistent with our results (see Table~\ref{fig:XCComparisonTable} for values).

\begin{figure}[H]
\centering
\includegraphics[width=0.75\textwidth]{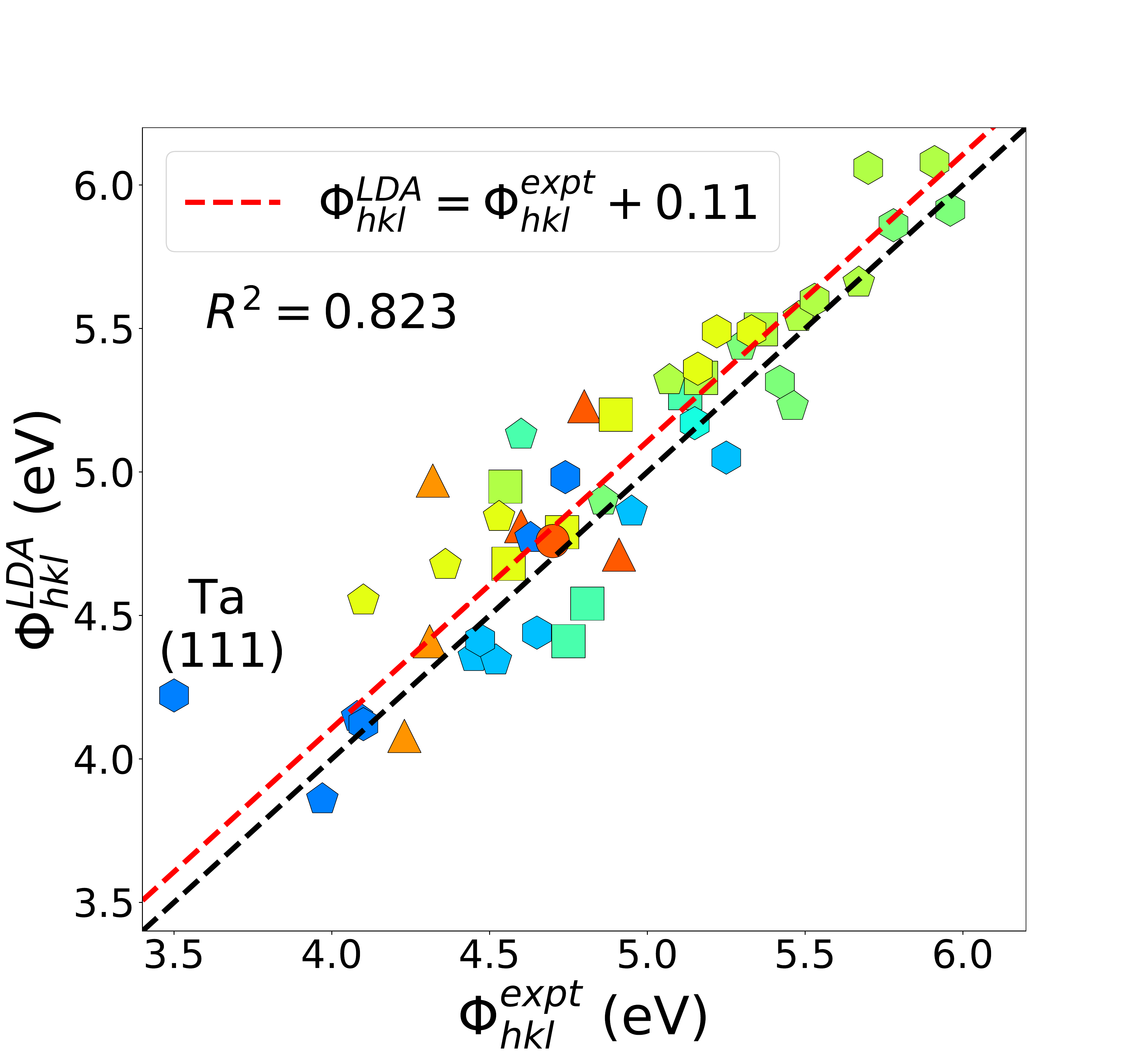}
\caption{\label{fig:all_lda_facets_vs_exp_facets} Plot of computed facet-dependent $\Phi_{\mbox{\scriptsize{hkl}}}$ using LDA from the literature~\cite{Waele2016, Ji2016, Patra2017} vs experimental values \cite{CRC, Kawano2008, Derry2015}.}
\end{figure}

Our values for $\Phi_{\mbox{\scriptsize{hkl}}}$ are in excellent agreement with those calculated using PBE, LDA and RPBE (as shown in Figure~\ref{fig:all_wf_facets_vs_gga_facets}, \ref{fig:all_wf_facets_vs_lda_facets} and \ref{fig:all_wf_facets_vs_rpbe_facets} respectively) with values of $R^2$ greater than 0.94 in all three cases. Unsurprisingly, there is smaller deviation when comparing our data to other GGA values (0.11 eV for PBE and 0.06 eV for rPBE) than to LDA values which are on average 0.38 eV higher. The major discrepancy between PBE and rPBE are for the (0001) surfaces of Y, Sc and Zn, with the PBE values being higher by 0.34 eV, 0.27 eV and 0.29 eV, respectively. Our values for $\Phi^{\mbox{\scriptsize{Sc}}}_{21\bar{3}1}$ and $\Phi^{\mbox{\scriptsize{Ba}}}_{\mbox{\scriptsize{210}}}$ are significantly lower when compared to PBE values obtained from the literature. Overall, the LDA-computed work functions are on average closer to the experimental values with a deviation of 0.11 eV (see Figure~\ref{fig:all_lda_facets_vs_exp_facets}).

\subsubsection{Work function of missing-row reconstructions}

\begin{figure}[H]
\center
\subfigure{\includegraphics[width=0.49\textwidth]{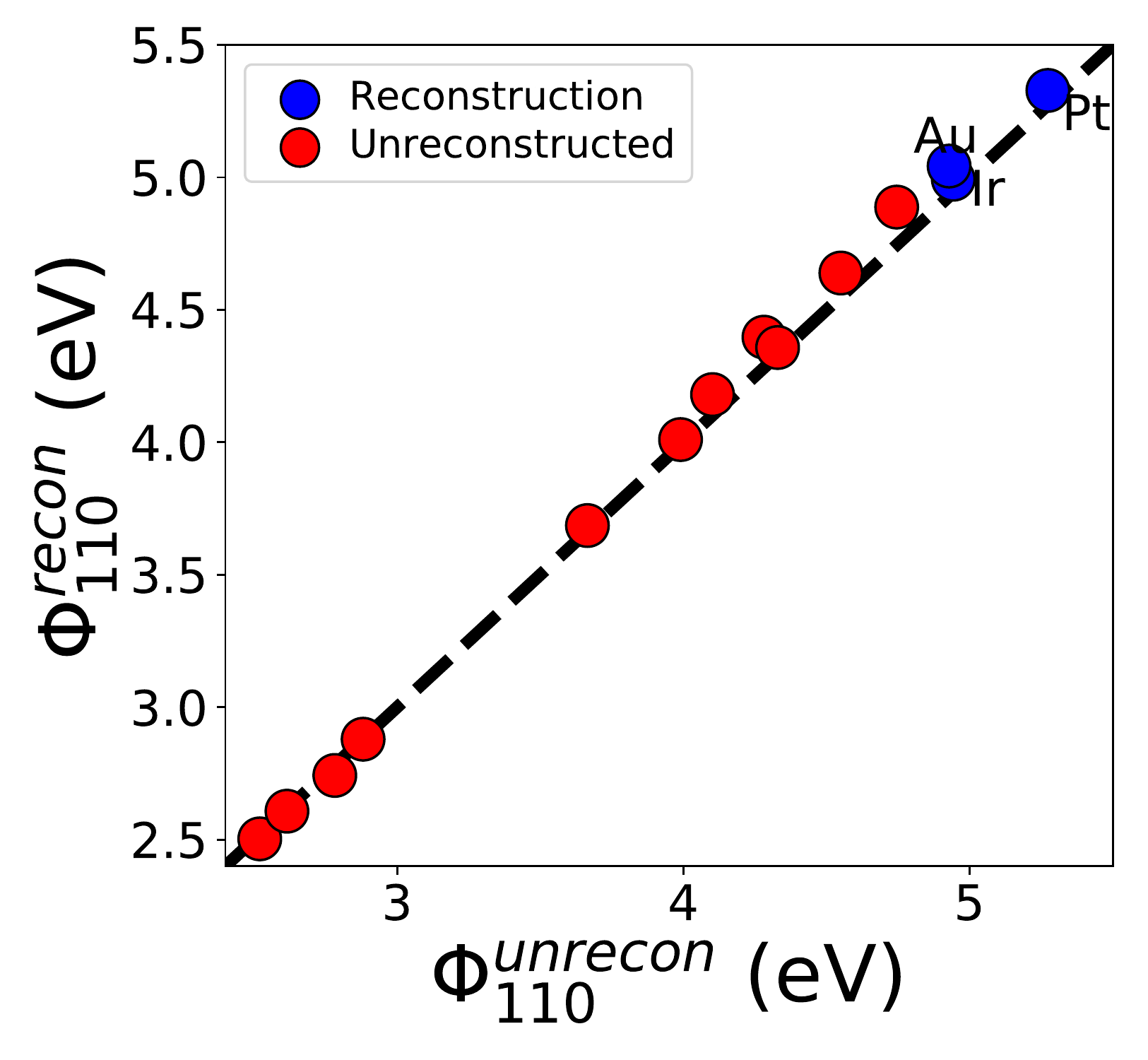}}
\caption{\label{fig:fcc110_reconstruction} Plot of the (110) work function for an unreconstructed ($\Phi^{\mbox{\scriptsize{unrecon}}}_{\mbox{\scriptsize{110}}}$) and $1\times2$ missing-row reconstructed ($\Phi^{\mbox{\scriptsize{recon}}}_{\mbox{\scriptsize{110}}}$) surface for fcc materials. Data points corresponding to materials where reconstruction is thermodynamically favorable ($-2$  meV$\mbox{\AA}^{-2} < \gamma^{\mbox{\scriptsize{recon}}}_{110} - \gamma^{\mbox{\scriptsize{unrecon}}}_{110}$) are labelled in blue.}
\end{figure}

Figure~\ref{fig:fcc110_reconstruction} compares the work function for the (110) missing-row reconstructed surface of face-centered cubic metals ($\Phi^{\mbox{\scriptsize{recon}}}_{\mbox{\scriptsize{110}}}$) to the work function of the corresponding unreconstructed surface ($\Phi^{\mbox{\scriptsize{unrecon}}}_{\mbox{\scriptsize{110}}}$). As found in our previous work, only Pt, Au and Ir have significantly lower surface energies for the (110) missing-row reconstruction compared to the unreconstructed surface, which is in agreement with experimental observations. In general, we find that reconstruction leads to a relatively small increase in the work functions, though the three fcc metals exhibiting a thermodynamic driving force to reconstruct also have the largest work functions. 

\section{Discussion}

\subsection{Periodic trends in the work function}

\begin{figure}[H]
\subfigure[Group number vs $\bar{\Phi}$ of transition metals.]
{\label{fig:wf_vs_group}\includegraphics[width=0.49\textwidth]{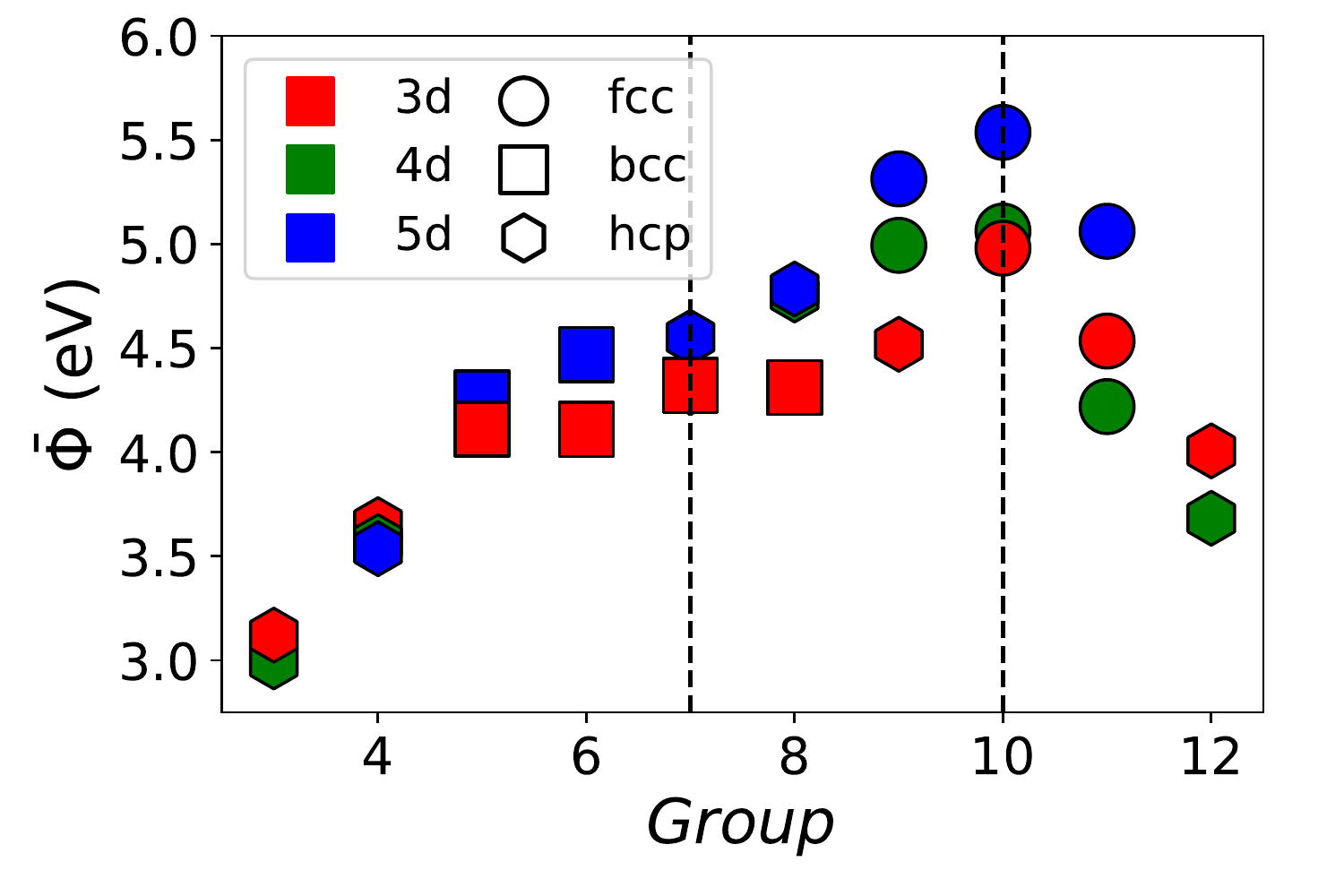}}
\subfigure[$\bar{\gamma}$ vs $\bar{\Phi}$ of transition metals.]
{\label{fig:wf_vs_se}\includegraphics[width=0.49\textwidth]{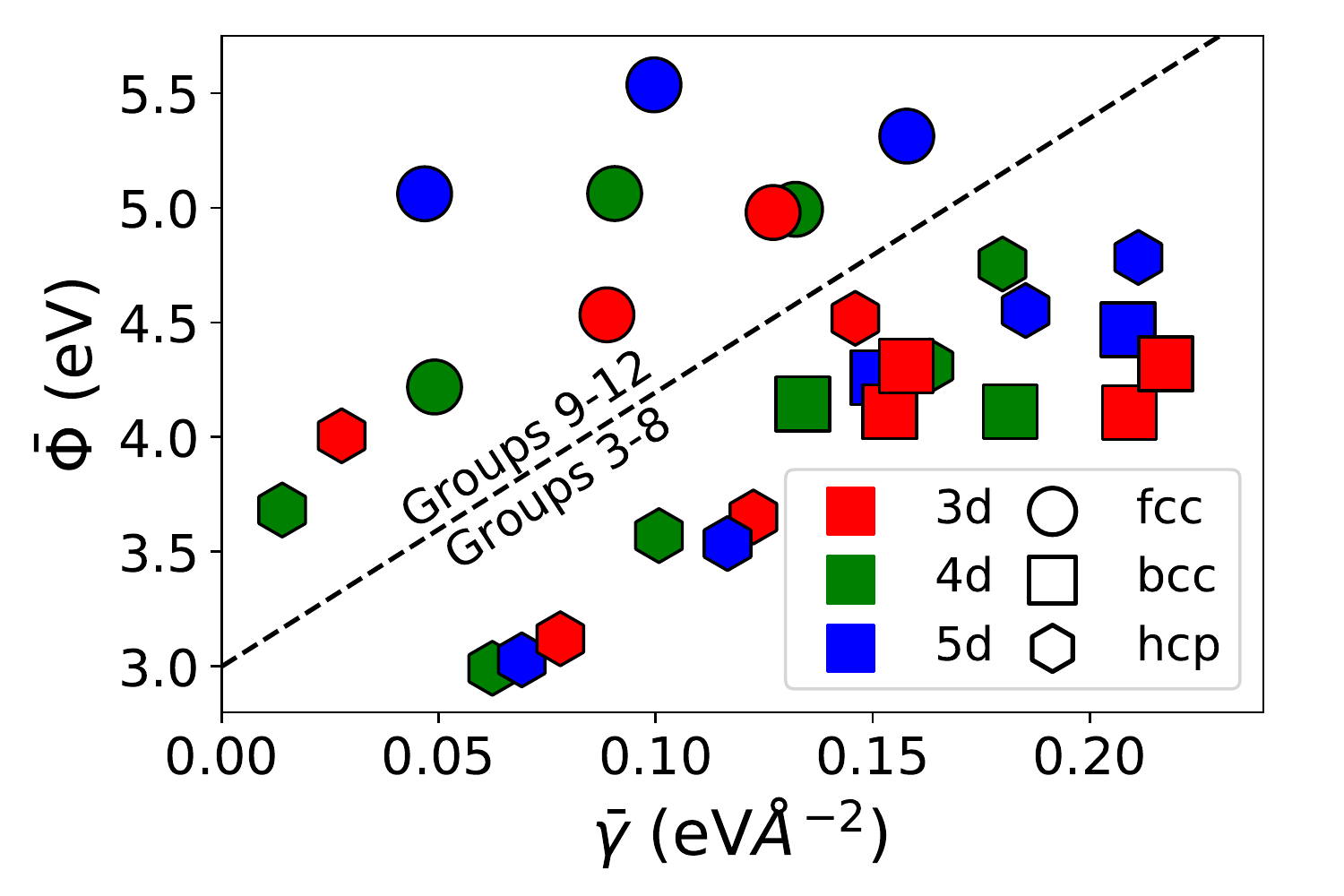}}
\subfigure[Group number vs $\bar{\Phi}$ of Lanthanides.]
{\label{fig:wf_vs_group_lanthanides}\includegraphics[width=0.49\textwidth]{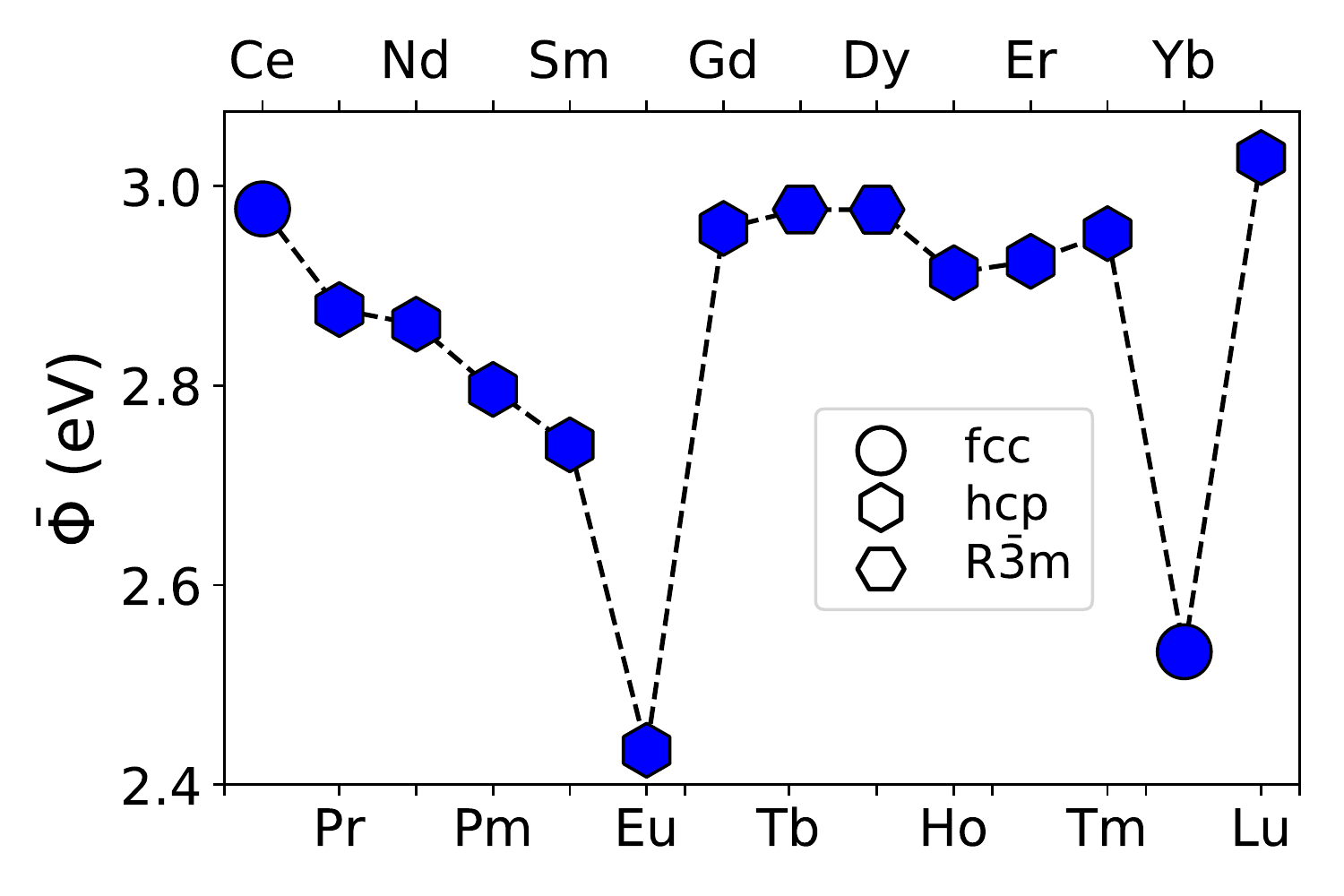}}
\caption{\label{fig:periodic_se_trends} Plot of $\bar{\Phi}$ versus (a) group number and (b) $\bar{\gamma}$ for transition metals, and (c) $\bar{\Phi}$ versus group number for lanthanides. The left and right dashed lines in (a) corresponds to the parabolic peak when plotting group number against $\bar{\gamma}$ and $\bar{\Phi}$ respectively.}
\end{figure}

Figure~\ref{fig:wf_vs_group} plots $\bar{\Phi}$ against the periodic group number for transition metals which demonstrates a parabolic behavior with the position of the parabolic maxima located at group 10 (Pt group). When plotting $\bar{\Phi}$ against the weighted surface energy ($\bar{\gamma}$) in Figure~\ref{fig:wf_vs_se}, we observe a split between elements above and below group 8. A similar parabolic trend when plotting $\bar{\gamma}$ against group number results in a maxima observed at group 7 rather than 10. The position of these parabolic peaks are related to the increasing cohesive energy resulting from the increasing number of half-filled $d$-orbitals as well as the width of the electronic $s$, $p$ and $d$ bands. For a more in-depth discussion, the interested reader is referred to the references herein\cite{Methfessel1992, Michaelides2010}. We further note that cohesive energy and thus surface energy are strongly correlated with mechanical properties which suggests a cohesive energy origin to previously observed trends between $\bar{\Phi}$ and mechanical properties of transition metals\cite{Rahemi2015, Hua2016, Lu2013}.

\begin{figure}[H]
\centering
\includegraphics[width=0.75\textwidth]{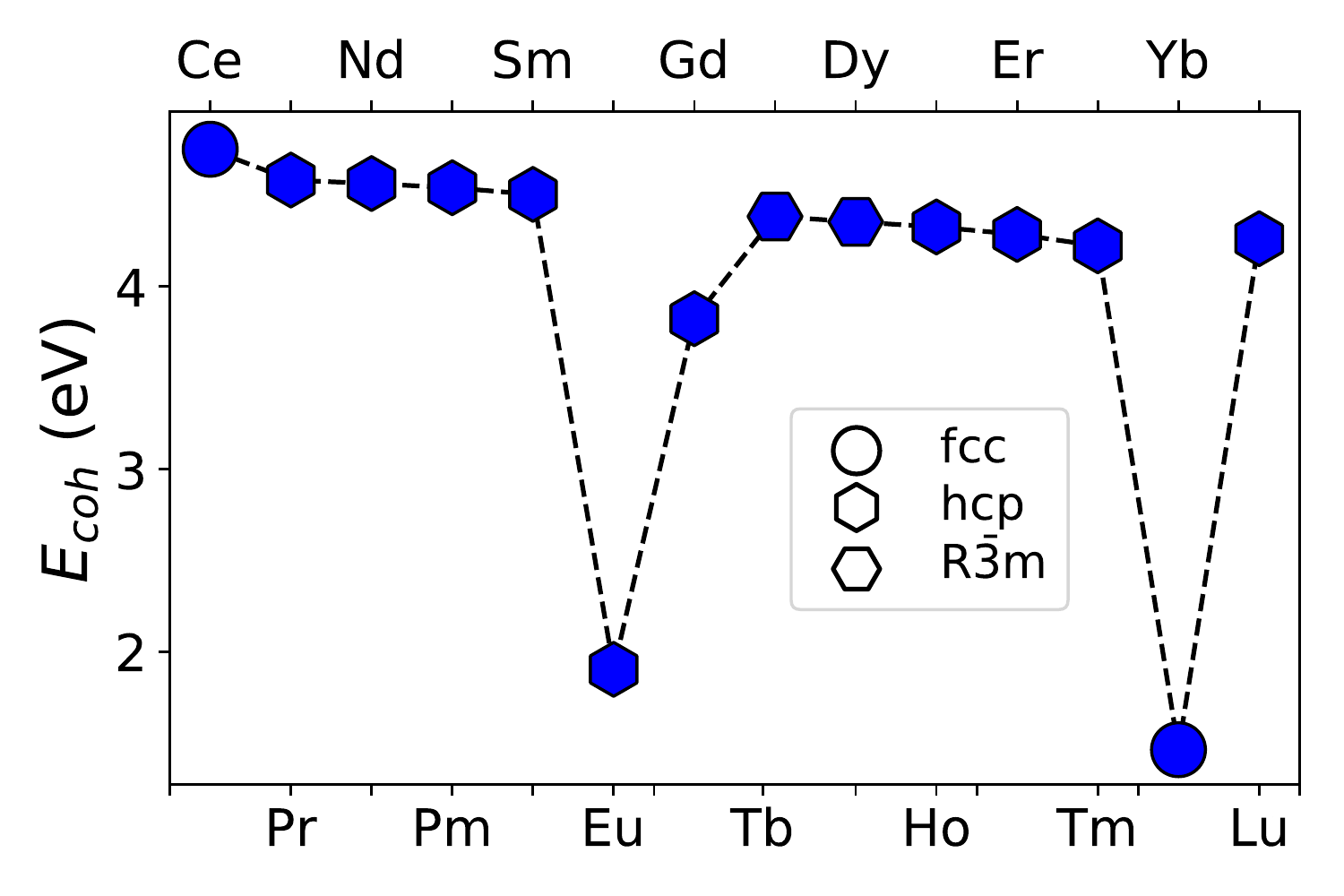}
\caption{\label{fig:ecoh_vs_group_lanthanides} Plot of the cohesive energy ($E_{coh}$) versus the group number for lanthanides.}
\end{figure}

To our knowledge, this work represents the first time DFT has been used to calculate the work functions of the lanthanides. Figure~\ref{fig:wf_vs_group_lanthanides} plots $\bar{\Phi}$ against the group number. A gradual decrease is observed for the half-filled lanthanides from Ce to Sm with a sharp decrease for Eu. The latter half of the lanthanides has a relatively constant value from Gd to Tm with a sharp decrease for Yb. Afterwards, a sharp increase is observed for Lu. These trends are consistent with trends of the cohesive energies of the lanthanides (see Figure~\ref{fig:ecoh_vs_group_lanthanides}), which in turn may be attributed to the gradual filling of the $4f$ orbitals. The two lowest $\bar{\Phi}$ are observed when the $4f$ orbitals are half filled (Eu) and completely filled (Yb), and these two elements also have the lowest cohesive energy and melting point among the lanthanides. With the exception of Ce, the computed $\bar{\Phi}$ underestimates the experimental values of $\Phi_{\mbox{\scriptsize{poly}}}^{\mbox{\scriptsize{expt}}}$ for the lanthanides with a standard deviation of 0.136 eV from Ce to Yb (see Table~\ref{fig:PolyAverageTable}).

\subsection{Discrepancies in the comparisons}

In general, our computed work functions are consistent with previous computational studies\cite{Patra2017, Waele2016, Wang2014c}. It is well known that the GGA(PBE) functional underestimates the intermediate range van der Waals (vdW) forces and Fermi energy while having no long-range vdW forces, which generally leads to an underestimated work function. Although LDA generally yields values closer to experiment than GGA(PBE), this agreement is due to the various errors inherent in LDA that work in tandem to provide an error cancellation.\cite{Patra2017} 

LDA/GGA are also known to have errors associated with overbinding/underbinding leading to smaller/larger cell volumes, lattice parameters and atomic distances, which can in turn influence surface properties. Larger atomic distances will decrease 2NN contributions to $\gamma_{\mbox{\scriptsize{hkl}}}$ and $\Phi_{\mbox{\scriptsize{hkl}}}$. This effect is especially prominent in the refractory metals Mo, Ta, Nb and W where the work function is shown to be dependent only on 1NN, but not 2NN (see Table~\ref{fig:SchmoluchoskiCorrelation}). Without 2NN contributions, the work function will be severely underestimated in these metals when compared to experimental values, thus explaining the increasing deviation for refractory metals. Furthermore, the surface energy scales well with 2NN for Nb and Ta (which have the greatest deviation in $\Phi_{\mbox{\scriptsize{hkl}}}$ of the four refractory metals), indicating that underbinding in GGA(PBE) is more consequential for work function than it is for surface energy.

Experimental error is also a potential source of discrepancies in our comparisons. $\bar{\Phi}^{\mbox{\scriptsize{As}}}$ is significantly higher while $\bar{\Phi}^{\mbox{\scriptsize{La}}}$ and $\bar{\Phi}^{\mbox{\scriptsize{Se}}}$ are significantly lower than that of their corresponding values for $\Phi_{\mbox{\scriptsize{poly}}}^{\mbox{\scriptsize{expt}}}$. Experimental values for these particular elements were taken from \citet{Michaelson1977} where surface contamination could lead to inaccuracies of up to 0.5 eV\cite{Michaelson1978} in the reported measurements. $\Phi^{\mbox{\scriptsize{As}}}_{\mbox{\scriptsize{poly}}}$ is also known to range from 3.75 to 5.4 eV which our value for $\bar{\Phi}^{\mbox{\scriptsize{As}}}$ lies between. Furthermore, the value of $\Phi^{\mbox{\scriptsize{Se}}}_{\mbox{\scriptsize{poly}}}$ was also determined using a photoelectric method which is known to yield erroneous values of work function for semiconductors. In addition, we opted to use the latest values available from the literature, which for $\Phi^{\mbox{\scriptsize{La}}}_{\mbox{\scriptsize{poly}}}$ came from~\citet{Michaelson1977}, despite measurements from~\citet{Rozkhov1971} being 0.54 eV closer to our DFT and previous linear muffin-tin orbital method values~\cite{Durakiewicz2001}.

Although the calculated $\Phi^{\mbox{\scriptsize{lowest}}}_{\mbox{\scriptsize{hkl}}}$ has previously been suggested as a good approximation of experimentally measured $\Phi_{\mbox{\scriptsize{poly}}}^{\mbox{\scriptsize{expt}}}$, we have shown that the Wulff-area-weighted $\bar{\Phi}$ provides a much closer estimate. This suggests that the eclipsing effect of lower work functions in PES signals of patchy surfaces may not be as prominent as once thought. Despite this, the values of $\Phi_{\mbox{\scriptsize{poly}}}^{\mbox{\scriptsize{expt}}}$ are higher than $\bar{\Phi}$ by an average value of 0.18 eV. \citet{Kawano2008} has argued that the weighted work function is more likely to follow a Boltzmann distribution with a higher value than that provided by Equation~\ref{eqn:average_work function}. In this context, temperature becomes an important factor in determining the work function\cite{Kiejna1979, Rahemi2015}. However, it is unaccounted for in our calculations which are assumed to take place at 0K.

\subsection{Effect of reconstruction on work function}

In general, the work function of the reconstructed facets are slightly larger than that of the unreconstructed facets. This can be due to the exposure of the \{111\} facets during reconstruction which generally have larger values of $\Phi_{\mbox{\scriptsize{hkl}}}$ than the (110) facet due to the lack of Smoluchowski smoothing in the flat \{111\} surfaces. It is not coincidental that the disparity between $\Phi^{\mbox{\scriptsize{recon}}}_{\mbox{\scriptsize{110}}}$ and $\Phi^{\mbox{\scriptsize{unrecon}}}_{\mbox{\scriptsize{110}}}$ for metals with surface reconstruction (Au, Ir, and Pt) is larger than for other metals. \citet{Ho1987a} previously explained that missing row reconstruction was the result of competing forces that contributed to the kinetic energy (KE) contributions to surface energy. A missing-row introduces additional broken $d$-bonds for transition metals which increases KE. At the same time, a larger surface area is created from the newly exposed \{111\} facets which will better facilitate the spreading of $s$ and $p$ electrons. This increase in electron spreading will lower the surface KE and for some elements such as Au, Ir and Pt, is enough to overcome the KE increase. Recall that electron spreading will increase work function, a tenet of the Smoluchowski model (see later section), which explains the larger increase in $\Phi^{\mbox{\scriptsize{recon}}}_{\mbox{\scriptsize{110}}}$ for reconstructed surfaces relative to $\Phi^{\mbox{\scriptsize{unrecon}}}_{\mbox{\scriptsize{110}}}$.

\subsection{Models for the work function}

\subsubsection{Smoluchowski rule}

\begin{figure}[H]
\subfigure[ $\Phi_{\mbox{\scriptsize{hkl}}}/\bar{\Phi}$ vs normalized broken bonds (fcc).]
{\label{fig:all_wf_per_a_fcc}\includegraphics[width=0.48\textwidth]{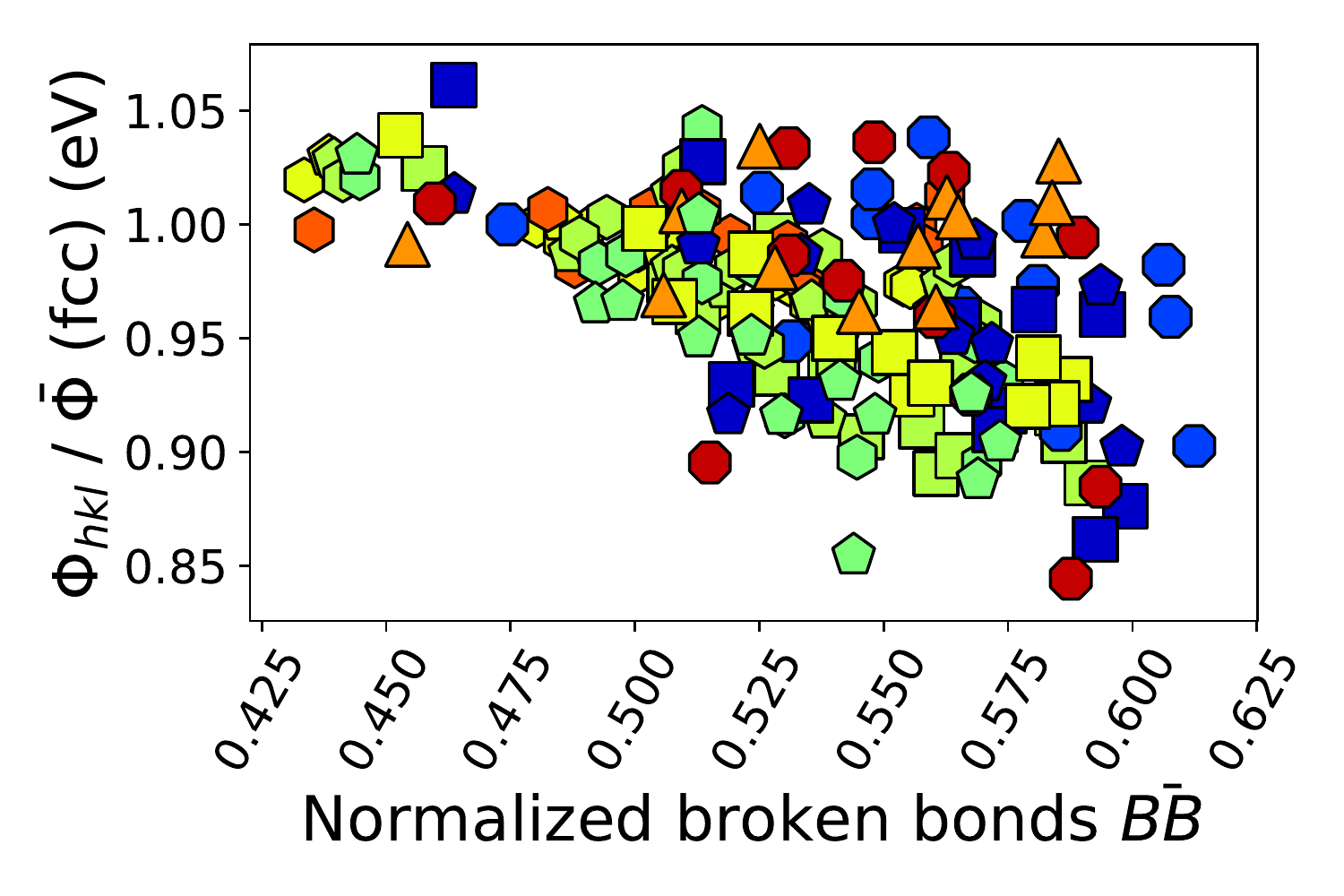}}
\subfigure[ $\Phi_{\mbox{\scriptsize{hkl}}}/\bar{\Phi}$ vs normalized broken bonds (bcc).]
{\label{fig:all_wf_per_a_bcc}\includegraphics[width=0.48\textwidth]{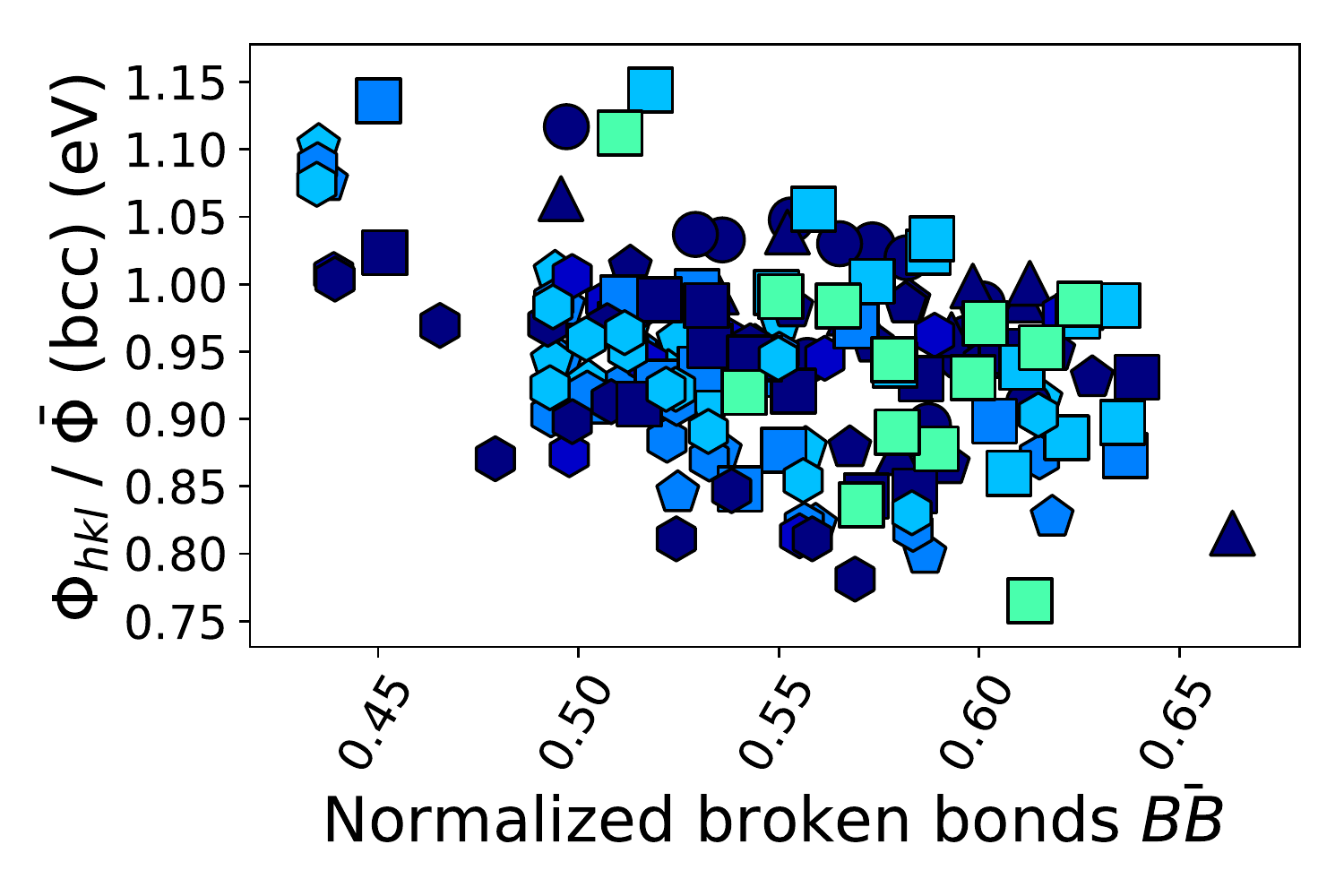}}
\quad
\subfigure[ $\Phi_{\mbox{\scriptsize{hkl}}}/\bar{\Phi}$ vs normalized broken bonds (hcp).]
{\label{fig:all_wf_per_a_hcp}\includegraphics[width=0.48\textwidth]{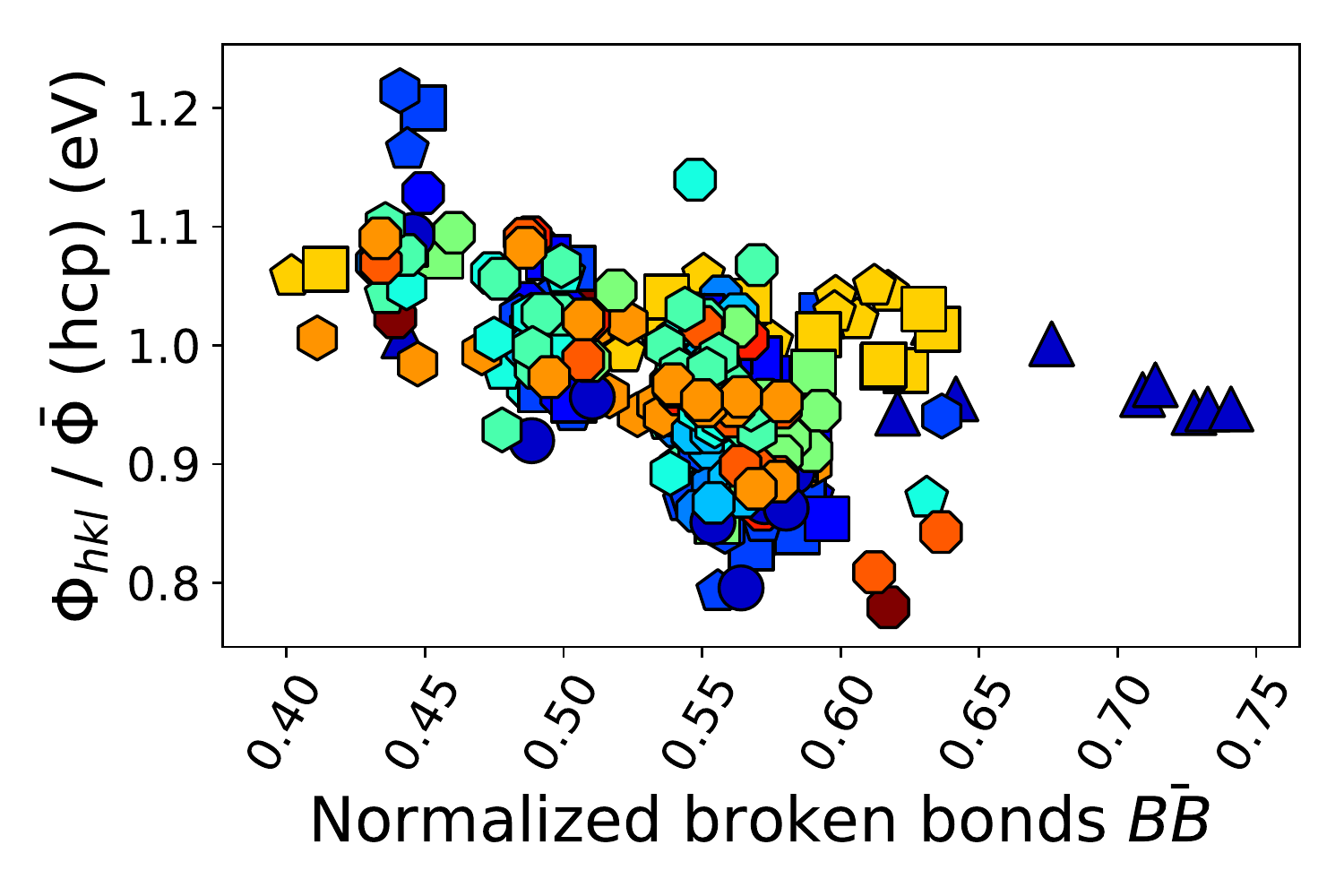}}
\subfigure{\includegraphics[width=0.48\textwidth]{periodic_legend}}
\caption{\label{fig:facet_vs_bb} Plot of the normalized broken bonds vs $\Phi_{\mbox{\scriptsize{hkl}}}/\bar{\Phi}$ normalized by average work function for (a) fcc, (b) bcc and (c) hcp structures for elemental crystalline solids commonly observed in literature. A legend indicating the element of each corresponding marker is shown at the bottom right.}
\end{figure}

Figure~\ref{fig:facet_vs_bb} plots $\Phi_{\mbox{\scriptsize{hkl}}}$ normalized by the average work function for each element $\bar{\Phi}$ as a function of the normalized broken bonds. All $r$ values obtained from comparing $\Phi_{\mbox{\scriptsize{hkl}}}$ and $\gamma_{\mbox{\scriptsize{hkl}}}$ to their respective normalized broken bonds and to each other are presented in Table~\ref{fig:SchmoluchoskiCorrelation}. Strong negative correlations were observed for the anisotropic work functions of 23 metals: Au, Ni, Ag, Pd, Cu, Rh, Nb, Mo, Li, Ta, W, Y, Lu, Ru, Zr, La, Tc, Sc, Tm, Re, Eu, Er and Ho. Among these systems, similar trends have been confirmed in previous studies for Ni, Cu, Ag, Mo and W, but not for Au and Pd~\cite{Wang2014c, Kawano2008, Waele2016}. Ta and Nb are the only metals where a stronger correlation with surface energy ($r_{\mbox{\scriptsize{Ta}}} = 0.96$, $r_{\mbox{\scriptsize{Nb}}} = 0.88$) is observed when modelling with 2NN and work function ($r_{\mbox{\scriptsize{Ta}}} = -0.78$, $r_{\mbox{\scriptsize{Nb}}} = -0.86$) when modelling with 1NN. A moderate negative correlation is observed for 20 metals: Pt, Ir, Ca, Sr, Na, V, Cs, K, Cr, Mg, Ti, Zn, Pr, Hf, Tl, Co, Be, Nd, Sm and Os. The remaining 9 metals have weak negative correlations with Cd and Al having no negative correlation ($r_{\mbox{\scriptsize{Cd}}}=0.01$, $r_{\mbox{\scriptsize{Al}}}=0.22$) at all.

\citet{Grepstad1976} has previously suggested that the Smoluchowski rule is valid only for systems with densely packed planes. It is well known that the c/a ratio of Cd is significantly larger than other hcp metals leading to sparsely packed planes along the (0001) direction~\cite{Gaston2010a} which can explain why Cd does not follow the Smoluchowski rule. However, although~\citet{Grepstad1976} was able to show that the computed values of $\Phi^{\mbox{\scriptsize{Cu}}}_{211}$  and $\Phi^{\mbox{\scriptsize{Al}}}_{\mbox{\scriptsize{hkl}}}$ are consistent with this explanation, our results clearly show that even for facets of Cu with MMI$>$1, bond breaking trends are still valid.

Alternatively, \citet{Fall1998} associates the anomalously low value of $\Phi^{\mbox{\scriptsize{Al}}}_{111}$ with the presence of $p$ orbitals parallel to its surface which are highly favored in electronically dense facets. By decreasing the valence electrons at the surface, the $p$ orbitals perpendicular to the surface become favored over the parallel $p$ orbitals. This leads to an increase in $\Phi^{\mbox{\scriptsize{Al}}}_{111}$ that will eventually lead to an anisotropy consistent with the Smoluchowski rule. It is possible that the same phenomenon can explain the lack of correlation in other $p$-block systems such as Pb ($r=-0.05$).

\subsection{An improved model for the work function of metals}

\begin{figure}[H]
\subfigure[$\bar{\Phi}$ vs $\chi$ and $R$]{\label{fig:X_and_R_vs_wf}\includegraphics[width=0.55\textwidth]{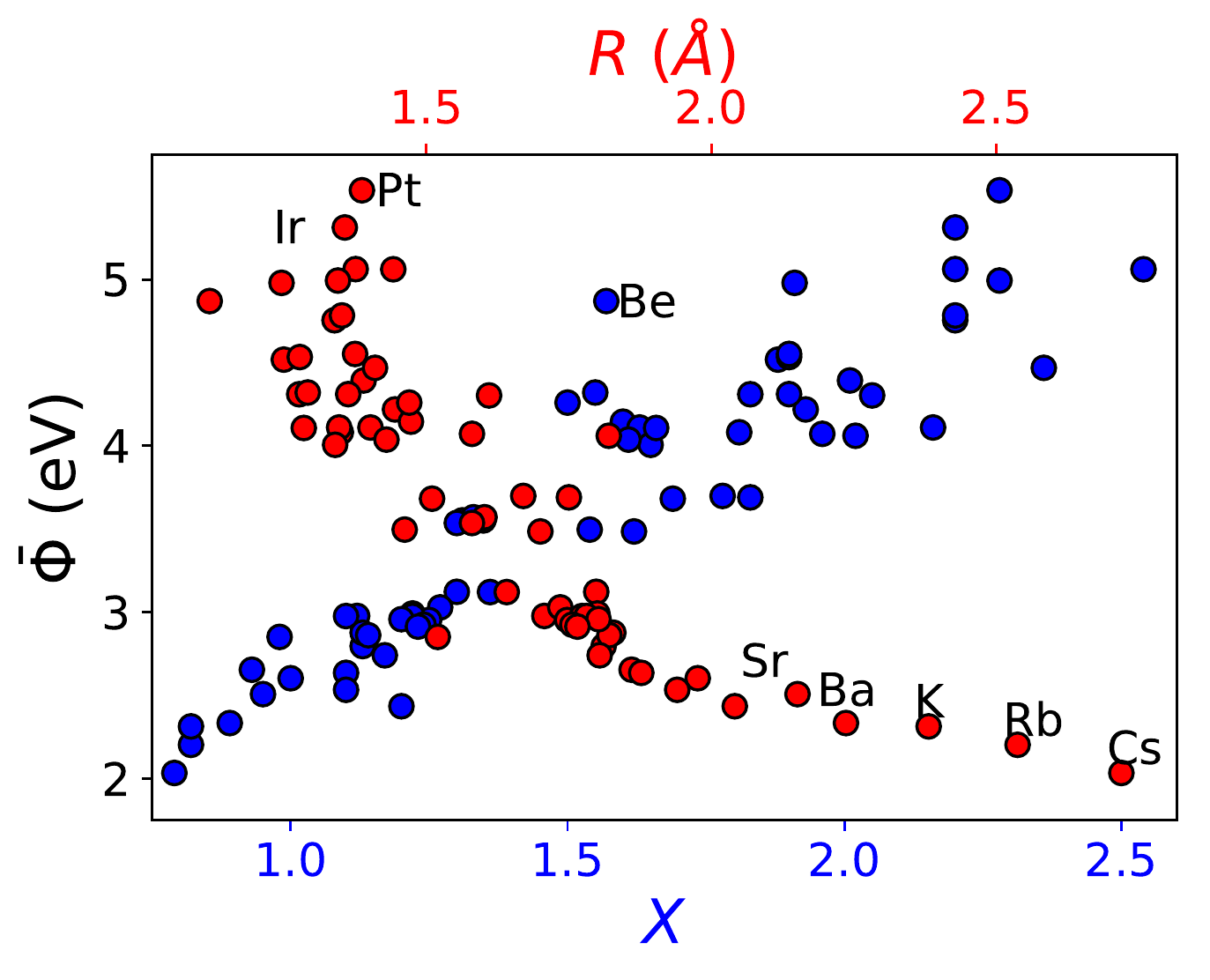}}
\subfigure[Improved model]{\label{fig:mlr_workfunction} \includegraphics[width=0.43\textwidth]{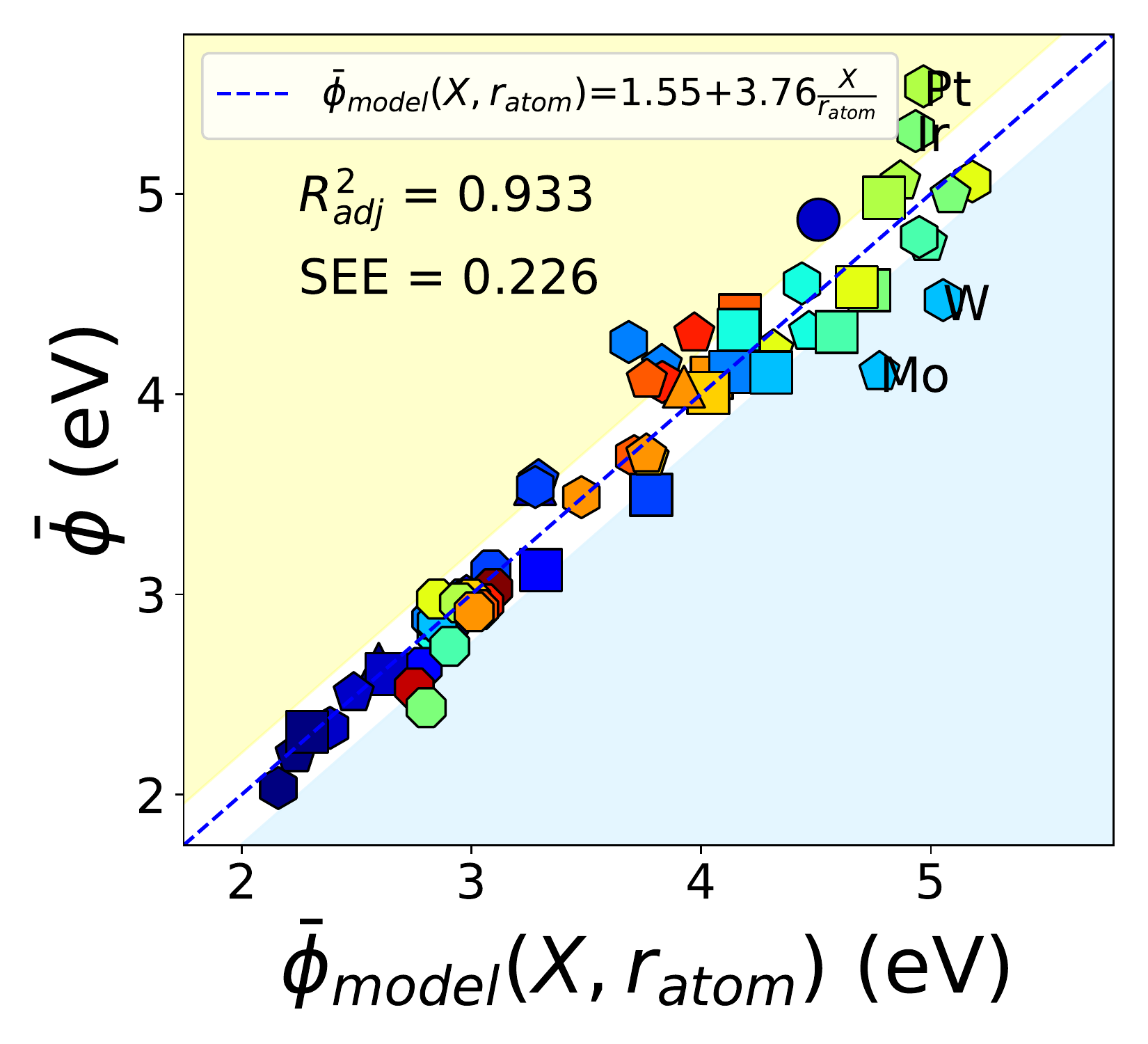}}
\caption{ Plot for the calculated $\bar{\Phi}$ against (a) Pauling electronegativity $\chi$ and metallic radius $R$, and (b) predictions from improved model for $\bar{\Phi} = 1.55 + 3.76 \frac{\chi}{r_{atom}}$, where $r_{atom} = ^3\sqrt{V_{atom}}$ and $V_{atom}$ is the unit cell volume per atom.}
\end{figure}

The comprehensive data set presented in this work affords us the ability to develop more robust models for the work function of the elemental metals. It has previously been well-established by \citet{Michaelson1978} and \citet{Miedema1973} that the work functions of metals have a positive linear relationship with the electronegativity $\chi$ of the metal. This may be explained by the fact that $\chi$ is a measure of how strongly electrons are bounded to the atom, and hence, the higher the $\chi$, the greater the energy needed to bring an electron from the bulk to the free vacuum ($\Phi$). Nevertheless, as can be seen from Figure \ref{fig:X_and_R_vs_wf}, it is clear that $\chi$ only explains $R^2 = 85.5\%$ of the variation in $\bar{\Phi}$ across the metals.

We carried an investigation of the relationship between $\bar{\Phi}$ and various atomic properties. As can be seen from Figure \ref{fig:X_and_R_vs_wf}, a strong, albeit non-linear, negative relationship is observed between $\bar{\Phi}$ and the metallic radius $R$. From Gauss' law, the potential inside an infinite charged plate is proportional to the bulk charge density times the square of the thickness of the plate, i.e., it scales charge per unit length of material. We postulate that the average work function is proportional to the electron density per unit length, similar in spirit to the traditional jellum work function model for metals~\cite{Jolla1971}. We performed a linear regression of $\bar{\Phi}$ against $\frac{\chi}{r_{atom}}$, where $\chi$ is related to the electron charge contributed per atom (in line with previous models) and $r_{atom} = ^3\sqrt{V_{atom}}$, where $V_{atom}$ is the unit cell volume per atom. As shown in Figure \ref{fig:mlr_workfunction}, this optimized model $\bar{\Phi} = 1.55 + 3.76 \frac{\chi}{r_{atom}}$ exhibits a much improved prediction accuracy for $\bar{\Phi}$, with a very high $R^2$ of 0.933 and a small SEE of 0.226 eV.

\section{Conclusion}

In conclusion we have constructed the largest database of anisotropic work functions to date. We have validated our database by comparing to both experimental and computational results from the literature and by confirming previously observed trends. In addition, we have also developed a technique for estimating the work function of a polycrystalline specimen using the Wulff shape and showed that it is a significantly more accurate estimate for experimental polycrystalline values than the lowest anisotropic work function. Using this large dataset, we have also extensively probed well-known empirical relationships for the work function, such as the Smoluchowski rule, and developed a substantially-improved prediction model for the work function of the metals from atomic properties such as the electronegativity and metallic radius.

\section*{Acknowledgement}

This work is supported by the Materials Project, funded by the U.S. Department of Energy, Office of Science, Office of Basic Energy Sciences, Materials Sciences and Engineering Division under Contract No. DE-AC02-05-CH11231: Materials Project program KC23MP. The authors also acknowledge computational resources provided by Triton Shared Computing Cluster (TSCC) at the University of California, San Diego, the National Energy Research Scientic Computing Centre (NERSC), and the Extreme Science and Engineering Discovery Environment (XSEDE) supported by National Science Foundation under Grant No. ACI-1053575.

\section*{Supplementary Information}

\newcommand{\Keywords}[1]{\par\noindent {\small{\em Keywords\/}: #1}}

\begin{center}
\begin{longtable}{ccccc}
\caption{The values of $\bar{\Phi}$ and  $\Phi^{\mbox{\scriptsize{lowest}}}_{\mbox{\scriptsize{hkl}}}$ (along with its corresponding Miller index) from high-throughput calculations and the experimental $\Phi_{\mbox{\scriptsize{poly}}}^{\mbox{\scriptsize{expt}}}$ from the literature.} \label{fig:PolyAverageTable}\\
\toprule
Material & $\bar{\Phi}$ (eV) & Surface & $\Phi^{\mbox{\scriptsize{lowest}}}_{\mbox{\scriptsize{hkl}}}$ (eV) & $\Phi_{\mbox{\scriptsize{poly}}}^{\mbox{\scriptsize{expt}}}$ (eV)\\
Li & 2.85 & (2, 2, 1) & 2.55 & 2.93~\cite{CRC}*\\
\hline
Al & 4.04 & (3, 3, 1) & 3.88 & 4.25~\cite{Kawano2008}*\\
\hline
Be & 4.87 & (1, 1, -2, 0) & 3.88 & 4.98~\cite{Michaelson1977}*\\
\hline
B & 4.67 & (1, 0, -1, 1) & 4.64 & 4.45~\cite{Michaelson1977}*\\
\hline
C & 4.24 & (0, 0, 1) & 4.22 & 4.65~\cite{Kawano2008}*\\
\hline
Na & 2.55 & (2, 1, 1) & 2.08 & 2.75~\cite{Michaelson1977}*\\
\hline
Mg & 3.55 & (2, 0, -2, 1) & 3.35 & 3.66~\cite{Michaelson1977}*\\
\hline
Si & 4.73 & (3, 1, 1) & 4.23 & 4.82~\cite{Kawano2008}*\\
\hline
K & 2.31 & (2, 1, 0) & 1.95 & 2.29~\cite{CRC}*\\
\hline
Ca & 2.77 & (3, 1, 0) & 2.39 & 2.87~\cite{Michaelson1977}*\\
\hline
Sc & 3.12 & (2, 0, -2, 1) & 2.66 & 3.50~\cite{Michaelson1977}*\\
\hline
Ti & 3.65 & (1, 1, -2, 0) & 3.03 & 4.33~\cite{Michaelson1977}*\\
\hline
V & 4.11 & (3, 2, 2) & 3.49 & 4.30~\cite{Michaelson1977}*\\
\hline
Cr & 4.11 & (3, 1, 1) & 3.53 & 4.50~\cite{Michaelson1977}*\\
\hline
Mn & 4.32 & (1, 1, 0) & 4.20 & 4.10~\cite{Michaelson1977}*\\
\hline
Co & 4.52 & (1, 0, -1, 2) & 3.85 & 5.00~\cite{Michaelson1977}*\\
\hline
Zn & 4.01 & (1, 0, -1, 1) & 3.91 & 4.33~\cite{Michaelson1977}*\\
\hline
Ga & 4.08 & (2, 1, 0) & 3.87 & 4.20~\cite{Michaelson1977}*\\
\hline
Ge & 4.39 & (3, 1, 0) & 4.05 & 5.00~\cite{Michaelson1977}*\\
\hline
As & 4.40 & (2, 2, -4, 1) & 4.31 & 3.75~\cite{Michaelson1977}*\\
\hline
Se & 5.22 & (0, 0, 1) & 5.07 & 5.90~\cite{Michaelson1977}*\\
\hline
Rb & 2.20 & (2, 1, 1) & 1.91 & 2.16~\cite{Michaelson1977}*\\
\hline
Sr & 2.51 & (2, 1, 0) & 2.26 & 2.59~\cite{Michaelson1977}*\\
\hline
Y & 2.99 & (2, 2, -4, 1) & 2.57 & 3.10~\cite{Michaelson1977}*\\
\hline
Zr & 3.57 & (1, 1, -2, 0) & 2.83 & 4.05~\cite{Michaelson1977}*\\
\hline
Ru & 4.75 & (1, 1, -2, 1) & 4.12 & 4.70~\cite{Kawano2008}*\\
\hline
Cd & 3.68 & (0, 0, 0, 1) & 3.63 & 4.08~\cite{CRC}*\\
\hline
In & 3.70 & (1, 0, 1) & 3.67 & 4.12~\cite{Michaelson1977}*\\
\hline
Sn & 4.07 & (3, 2, 1) & 3.87 & 4.42~\cite{Michaelson1977}*\\
\hline
Sb & 4.30 & (2, 2, -4, 1) & 4.20 & 4.55~\cite{Michaelson1977}*\\
\hline
Te & 4.51 & (1, 0, -1, 1) & 4.29 & 4.95~\cite{Michaelson1977}*\\
\hline
Cs & 2.03 & (1, 1, 1) & 1.59 & 1.95~\cite{CRC}*\\
\hline
Ba & 2.33 & (2, 1, 0) & 1.90 & 2.53~\cite{CRC}*\\
\hline
La & 2.64 & (2, 2, -4, 1) & 2.42 & 3.50~\cite{Michaelson1977}*\\
\hline
Hf & 3.54 & (1, 1, -2, 0) & 2.98 & 3.90~\cite{Michaelson1977}*\\
\hline
W & 4.47 & (3, 1, 0) & 3.71 & 4.55~\cite{Kawano2008}*\\
\hline
Re & 4.55 & (1, 1, -2, 1) & 4.06 & 4.95~\cite{Kawano2008}*\\
\hline
Os & 4.78 & (1, 1, -2, 1) & 4.44 & 4.84~\cite{Kawano2008}*\\
\hline
Tl & 3.47 & (1, 0, -1, 1) & 3.13 & 3.84~\cite{Michaelson1977}*\\
\hline
Pb & 3.69 & (1, 1, 0) & 3.60 & 4.25~\cite{Michaelson1977}*\\
\hline
Bi & 4.06 & (2, 2, -4, 1) & 3.93 & 4.22~\cite{Michaelson1977}*\\
\hline
Th & 3.12 & (1, 1, 1) & 2.99 & 3.40~\cite{Michaelson1977}*\\
\hline
Ce & 2.98 & (2, 1, 0) & 2.69 & 2.90~\cite{Michaelson1977}*\\
\hline
Nd & 2.86 & (2, 2, -4, 1) & 2.48 & 3.20~\cite{Michaelson1977}*\\
\hline
Sm & 2.74 & (1, 1, -2, 1) & 2.54 & 2.70~\cite{Michaelson1977}*\\
\hline
Eu & 2.43 & (1, 0, -1, 0) & 2.08 & 2.50~\cite{Michaelson1977}*\\
\hline
Gd & 2.96 & (2, 1, -3, 0) & 2.38 & 3.10~\cite{Michaelson1977}*\\
\hline
Tb & 2.98 & (2, -1, -1, 2) & 2.53 & 3.00~\cite{Michaelson1977}*\\
\hline
Lu & 3.03 & (2, 1, -3, 0) & 2.36 & 3.30~\cite{Michaelson1977}*\\
\hline
Fe & 4.31 & (3, 2, 2) & 3.30 & 4.36~\cite{Kawano2008}*\\
\hline
Ni & 4.98 & (2, 1, 0) & 4.41 & 5.02~\cite{Kawano2008}*\\
\hline
Cu & 4.53 & (3, 1, 0) & 4.17 & 4.55~\cite{Kawano2008}*\\
\hline
Nb & 4.15 & (3, 1, 0) & 3.31 & 3.99~\cite{Kawano2008}*\\
\hline
Mo & 4.11 & (3, 1, 0) & 3.53 & 4.39~\cite{Kawano2008}*\\
\hline
Rh & 4.99 & (1, 1, 0) & 4.27 & 4.88~\cite{Kawano2008}*\\
\hline
Pd & 5.06 & (1, 1, 0) & 4.63 & 5.33~\cite{Kawano2008}*\\
\hline
Ag & 4.22 & (2, 1, 1) & 3.97 & 4.33~\cite{Kawano2008}*\\
\hline
Ta & 4.26 & (3, 1, 0) & 3.49 & 4.25~\cite{Kawano2008}*\\
\hline
Ir & 5.31 & (3, 2, 0) & 4.75 & 5.27~\cite{Kawano2008}*\\
\hline
Pt & 5.54 & (1, 1, 0) & 5.19 & 5.41~\cite{Kawano2008}*\\
\hline
Au & 5.06 & (2, 1, 0) & 4.71 & 5.20~\cite{Kawano2008}*\\
\hline
Pr & 2.88 & (2, 2, -4, 1) & 2.48 & 2.96~\cite{Rozkhov1971}*\\
\hline
Dy & 2.98 & (2, -1, -1, 2) & 2.55 & 3.25~\cite{Rozkhov1971}*\\
\hline
Ho & 2.91 & (2, 1, -3, 1) & 2.56 & 3.22~\cite{Rozkhov1971}*\\
\hline
Er & 2.92 & (2, 1, -3, 0) & 2.37 & 3.25~\cite{Rozkhov1971}*\\
\hline
Tm & 2.95 & (2, 1, -3, 1) & 2.54 & 3.10~\cite{Rozkhov1971}*\\
\hline

\end{longtable}
\end{center}

\begin{minipage}{1\textwidth}
{\footnotesize
* See reference herein
}
\end{minipage}

\clearpage

\begin{center}
\begin{longtable}{ccccccc}
\caption{A comparison of the high-throughput values to experimental and computed values for materials from the literature.} \label{fig:XCComparisonTable}\\

\hline Material & Surface & \multicolumn{5}{c}{Work function $\Phi$ ($eV$)} \\
& & HT & GGA(RPBE) & GGA(PBE) & LDA & Experiment \\
Ru & ($0001$) & 4.96 & 4.83 & 4.97~\cite{Ji2016} & 5.31~\cite{Ji2016} & \\
\hline
 & ($10\overline{1}0$) & 4.7 &  & 4.79~\cite{Ji2016} & 5.13~\cite{Ji2016} & 4.6~\cite{Derry2015}*\\
\hline
 & ($10\overline{1}1$) & 4.84 &  & 4.91~\cite{Ji2016} & 5.26~\cite{Ji2016} & \\
\hline
 & ($10\overline{1}2$) & 4.45 &  & 4.5~\cite{Ji2016} & 4.85~\cite{Ji2016} & \\
\hline
 & ($11\overline{2}1$) & 4.12 &  & 4.39~\cite{Ji2016} & 4.76~\cite{Ji2016} & \\
\hline
 & ($21\overline{3}0$) & 4.28 &  & 4.47~\cite{Ji2016} & 4.86~\cite{Ji2016} & \\
\hline
Pt & ($100$) & 5.68 & 5.47 & 5.69~\cite{Patra2017} & 6.06~\cite{Waele2016} & 5.7~\cite{Derry2015}*\\
\hline
 & ($111$) & 5.64 & 5.72 & 5.12~\cite{Patra2017} & 6.08~\cite{Waele2016} & 5.91~\cite{Derry2015}*\\
\hline
 & ($110$) & 5.33 & 5.25 & 4.94~\cite{Patra2017} & 5.6~\cite{Waele2016} & 5.53~\cite{Derry2015}*\\
\hline
 & ($321$) & 5.34 &  & 5.44~\cite{Wang2014c} &  & 5.4~\cite{Kawano2008}*\\
\hline
 & ($211$) & 5.43 &  & 5.55~\cite{Wang2014c} &  & \\
\hline
 & ($310$) & 5.44 &  & 5.42~\cite{Wang2014c} &  & \\
\hline
 & ($210$) & 5.3 &  &  &  & 5.18~\cite{Kawano2008}*\\
\hline
 & ($311$) & 5.47 &  &  &  & 5.5~\cite{Kawano2008}*\\
\hline
 & ($320$) & 5.2 &  &  &  & 5.2~\cite{Kawano2008}*\\
\hline
 & ($331$) & 5.24 &  &  &  & 5.12~\cite{Kawano2008}*\\
\hline
Ni & ($100$) & 4.95 & 4.71 & 4.9~\cite{Waele2016} & 5.33~\cite{Waele2016} & 5.17~\cite{Derry2015}*\\
\hline
 & ($111$) & 5.1 & 4.89 & 5.02~\cite{Waele2016} & 5.5~\cite{Waele2016} & 5.36~\cite{Derry2015}*\\
\hline
 & ($110$) & 4.43 & 4.26 & 4.49~\cite{Waele2016} & 4.95~\cite{Waele2016} & 4.55~\cite{Derry2015}*\\
\hline
Nb & ($100$) & 3.43 & 3.42 & 3.55~\cite{Wang2014c} & 3.86~\cite{Waele2016} & 3.97~\cite{Derry2015}*\\
\hline
 & ($110$) & 4.46 & 4.43 & 4.49~\cite{Wang2014c} & 4.77~\cite{Waele2016} & 4.63~\cite{Derry2015}*\\
\hline
 & ($111$) & 3.63 &  & 3.77~\cite{Wang2014c} & 4.15~\cite{Waele2016} & 4.08~\cite{Derry2015}*\\
\hline
 & ($210$) & 3.86 &  & 3.97~\cite{Wang2014c} &  & \\
\hline
 & ($331$) & 4.08 &  & 4.15~\cite{Wang2014c} &  & \\
\hline
 & ($311$) & 3.4 &  & 3.64~\cite{Wang2014c} &  & 4.29~\cite{Kawano2008}*\\
\hline
 & ($310$) & 3.31 &  &  &  & 4.18~\cite{Kawano2008}*\\
\hline
 & ($211$) & 3.77 &  &  &  & 4.45~\cite{Kawano2008}*\\
\hline
Y & ($0001$) & 3.18 &  & 3.18~\cite{Waele2016} & 3.44~\cite{Waele2016} & \\
\hline
 & ($10\overline{1}0$) & 3.23 &  & 3.31~\cite{Waele2016} & 3.47~\cite{Waele2016} & \\
\hline
 & ($21\overline{3}0$) & 2.76 &  & 3.0~\cite{Waele2016} & 3.2~\cite{Waele2016} & \\
\hline
Pd & ($111$) & 5.21 & 4.94 & 5.32~\cite{Patra2017} & 5.66~\cite{Patra2017} & 5.67~\cite{Derry2015}*\\
\hline
 & ($110$) & 4.63 & 4.56 & 4.95~\cite{Patra2017} & 5.32~\cite{Patra2017} & 5.07~\cite{Derry2015}*\\
\hline
 & ($100$) & 5.13 &  & 5.12~\cite{Patra2017} & 5.54~\cite{Patra2017} & 5.48~\cite{Derry2015}*\\
\hline
 & ($321$) & 4.86 &  & 4.89~\cite{Wang2014c} &  & \\
\hline
 & ($211$) & 4.92 &  & 4.99~\cite{Wang2014c} &  & \\
\hline
 & ($310$) & 4.93 &  & 4.86~\cite{Wang2014c} &  & \\
\hline
Rh & ($100$) & 5.01 & 4.9 & 5.04~\cite{Patra2017} & 5.44~\cite{Patra2017} & 5.3~\cite{Derry2015}*\\
\hline
 & ($111$) & 5.15 & 4.77 & 5.0~\cite{Patra2017} & 5.23~\cite{Patra2017} & 5.46~\cite{Derry2015}*\\
\hline
 & ($110$) & 4.27 & 4.25 & 4.53~\cite{Patra2017} & 4.9~\cite{Patra2017} & 4.86~\cite{Derry2015}*\\
\hline
 & ($321$) & 4.57 &  & 4.65~\cite{Wang2014c} &  & \\
\hline
 & ($211$) & 4.75 &  & 4.87~\cite{Wang2014c} &  & \\
\hline
 & ($310$) & 4.62 &  & 4.74~\cite{Wang2014c} &  & \\
\hline
Tc & ($0001$) & 4.63 & 4.47 & 4.69~\cite{Ji2016} & 4.95~\cite{Ji2016} & \\
\hline
 & ($10\overline{1}0$) & 4.22 &  & 4.5~\cite{Ji2016} & 4.83~\cite{Ji2016} & \\
\hline
 & ($10\overline{1}1$) & 3.76 &  & 4.7~\cite{Ji2016} & 5.05~\cite{Ji2016} & \\
\hline
 & ($10\overline{1}2$) & 4.35 &  & 4.31~\cite{Ji2016} & 4.67~\cite{Ji2016} & \\
\hline
 & ($11\overline{2}1$) & 3.8 &  & 4.09~\cite{Ji2016} & 4.44~\cite{Ji2016} & \\
\hline
 & ($2\overline{1}\overline{1}2$) & 4.19 &  & 4.28~\cite{Ji2016} & 4.63~\cite{Ji2016} & \\
\hline
 & ($21\overline{3}0$) & 4.05 &  & 4.26~\cite{Ji2016} & 4.58~\cite{Ji2016} & \\
\hline
Ta & ($100$) & 3.71 & 3.67 & 4.1~\cite{Wang2014c} & 4.12~\cite{Waele2016} & 4.1~\cite{Derry2015}*\\
\hline
 & ($110$) & 4.64 & 4.64 & 4.96~\cite{Wang2014c} & 4.98~\cite{Waele2016} & 4.74~\cite{Derry2015}*\\
\hline
 & ($111$) & 3.71 &  & 4.2~\cite{Wang2014c} & 4.22~\cite{Waele2016} & 3.5~\cite{Derry2015}*\\
\hline
 & ($210$) & 4.04 &  & 4.34~\cite{Wang2014c} &  & \\
\hline
 & ($331$) & 4.21 &  & 4.62~\cite{Wang2014c} &  & \\
\hline
 & ($311$) & 3.5 &  & 4.1~\cite{Wang2014c} &  & \\
\hline
 & ($211$) & 3.92 &  &  &  & 4.45~\cite{Kawano2008}*\\
\hline
Fe & ($110$) & 4.79 & 4.66 & 4.74~\cite{Waele2016} & 5.28~\cite{Waele2016} & 5.12~\cite{Derry2015}*\\
\hline
 & ($111$) & 4.25 &  & 3.86~\cite{Waele2016} & 4.54~\cite{Waele2016} & 4.81~\cite{Derry2015}*\\
\hline
 & ($100$) & 3.96 &  & 3.89~\cite{Waele2016} & 4.41~\cite{Waele2016} & 4.75~\cite{Derry2015}*\\
\hline
Hf & ($0001$) & 4.29 & 4.12 & 4.34~\cite{Waele2016} & 4.58~\cite{Waele2016} & \\
\hline
 & ($10\overline{1}0$) & 3.62 &  & 3.94~\cite{Waele2016} & 4.21~\cite{Waele2016} & \\
\hline
 & ($21\overline{3}0$) & 3.23 &  & 3.12~\cite{Waele2016} & 3.38~\cite{Waele2016} & \\
\hline
Mo & ($100$) & 3.76 & 3.61 & 3.84~\cite{Wang2014c} & 4.36~\cite{Waele2016} & 4.45~\cite{Derry2015}*\\
\hline
 & ($110$) & 4.53 & 4.46 & 4.51~\cite{Wang2014c} & 4.86~\cite{Waele2016} & 4.95~\cite{Derry2015}*\\
\hline
 & ($111$) & 3.76 &  & 3.94~\cite{Wang2014c} & 4.34~\cite{Waele2016} & 4.52~\cite{Derry2015}*\\
\hline
 & ($210$) & 4.0 &  & 4.11~\cite{Wang2014c} &  & \\
\hline
 & ($331$) & 4.15 &  & 4.25~\cite{Wang2014c} &  & \\
\hline
 & ($311$) & 3.61 &  & 3.89~\cite{Wang2014c} &  & \\
\hline
 & ($211$) & 3.81 &  &  &  & 4.36~\cite{Kawano2008}*\\
\hline
W & ($100$) & 4.04 & 4.03 & 4.09~\cite{Wang2014c} & 4.44~\cite{Waele2016} & 4.65~\cite{Derry2015}*\\
\hline
 & ($110$) & 4.8 & 4.64 & 4.76~\cite{Wang2014c} & 5.05~\cite{Waele2016} & 5.25~\cite{Derry2015}*\\
\hline
 & ($111$) & 3.98 &  & 4.24~\cite{Wang2014c} & 4.41~\cite{Waele2016} & 4.47~\cite{Derry2015}*\\
\hline
 & ($210$) & 4.22 &  & 4.29~\cite{Wang2014c} &  & 4.38~\cite{Kawano2008}*\\
\hline
 & ($331$) & 4.39 &  & 4.46~\cite{Wang2014c} &  & \\
\hline
 & ($311$) & 3.82 &  & 4.1~\cite{Wang2014c} &  & 4.46~\cite{Kawano2008}*\\
\hline
 & ($211$) & 4.29 &  &  &  & 4.76~\cite{Derry2015}*\\
\hline
 & ($321$) & 4.13 &  &  &  & 4.49~\cite{Kawano2008}*\\
\hline
 & ($310$) & 3.71 &  &  &  & 4.32~\cite{Kawano2008}*\\
\hline
V & ($100$) & 3.59 & 3.56 & 3.74~\cite{Waele2016} & 4.07~\cite{Waele2016} & \\
\hline
 & ($110$) & 4.67 & 4.66 & 4.74~\cite{Waele2016} & 5.02~\cite{Waele2016} & \\
\hline
 & ($111$) & 3.6 &  & 3.87~\cite{Waele2016} & 4.23~\cite{Waele2016} & \\
\hline
Sc & ($0001$) & 3.39 & 3.66 & 3.33~\cite{Waele2016} & 3.56~\cite{Waele2016} & \\
\hline
 & ($10\overline{1}0$) & 3.35 &  & 3.56~\cite{Waele2016} & 3.77~\cite{Waele2016} & \\
\hline
 & ($21\overline{3}0$) & 3.09 &  & 3.18~\cite{Waele2016} & 3.35~\cite{Waele2016} & \\
\hline
 & ($21\overline{3}1$) & 2.84 &  & 3.42~\cite{Waele2016} & 3.39~\cite{Waele2016} & \\
\hline
Os & ($0001$) & 5.27 & 5.1 & 5.32~\cite{Ji2016} & 5.64~\cite{Ji2016} & \\
\hline
 & ($10\overline{1}0$) & 4.44 &  & 5.17~\cite{Ji2016} & 5.45~\cite{Ji2016} & \\
\hline
 & ($10\overline{1}1$) & 4.77 &  & 5.23~\cite{Ji2016} & 5.53~\cite{Ji2016} & \\
\hline
 & ($10\overline{1}2$) & 4.78 &  & 4.85~\cite{Ji2016} & 5.15~\cite{Ji2016} & \\
\hline
 & ($11\overline{2}1$) & 4.44 &  & 4.67~\cite{Ji2016} & 4.98~\cite{Ji2016} & \\
\hline
 & ($2\overline{1}\overline{1}2$) & 4.93 &  & 4.9~\cite{Ji2016} & 5.22~\cite{Ji2016} & \\
\hline
 & ($21\overline{3}0$) & 4.68 &  & 4.87~\cite{Ji2016} & 5.2~\cite{Ji2016} & \\
\hline
Zn & ($0001$) & 3.92 & 4.21 & 4.08~\cite{Waele2016} & 4.44~\cite{Waele2016} & \\
\hline
 & ($10\overline{1}0$) & 4.08 &  & 4.33~\cite{Waele2016} & 4.7~\cite{Waele2016} & \\
\hline
 & ($21\overline{3}0$) & 3.92 &  & 4.03~\cite{Waele2016} & 4.3~\cite{Waele2016} & \\
\hline
 & ($21\overline{3}1$) & 3.93 &  & 4.13~\cite{Waele2016} & 4.43~\cite{Waele2016} & \\
\hline
Co & ($0001$) & 4.86 & 4.75 & 4.92~\cite{Waele2016} & 5.39~\cite{Waele2016} & \\
\hline
 & ($10\overline{1}0$) & 4.55 &  & 4.7~\cite{Waele2016} & 5.14~\cite{Waele2016} & \\
\hline
 & ($21\overline{3}0$) & 4.37 &  & 4.39~\cite{Waele2016} & 4.81~\cite{Waele2016} & \\
\hline
 & ($21\overline{3}1$) & 4.34 &  & 4.33~\cite{Waele2016} & 4.76~\cite{Waele2016} & \\
\hline
Ag & ($100$) & 4.21 & 3.96 & 4.26~\cite{Patra2017} & 4.68~\cite{Waele2016} & 4.36~\cite{Derry2015}*\\
\hline
 & ($111$) & 4.34 & 4.21 & 4.49~\cite{Patra2017} & 4.84~\cite{Waele2016} & 4.53~\cite{Derry2015}*\\
\hline
 & ($110$) & 4.05 & 4.04 & 4.16~\cite{Patra2017} & 4.55~\cite{Waele2016} & 4.1~\cite{Derry2015}*\\
\hline
 & ($321$) & 4.01 &  & 4.14~\cite{Wang2014c} &  & \\
\hline
 & ($211$) & 3.97 &  & 4.24~\cite{Wang2014c} &  & \\
\hline
 & ($310$) & 4.06 &  & 4.08~\cite{Wang2014c} &  & \\
\hline
Re & ($0001$) & 4.77 & 4.85 & 4.88~\cite{Ji2016} & 5.17~\cite{Ji2016} & 5.15~\cite{Kawano2008}*\\
\hline
 & ($10\overline{1}0$) & 4.58 &  & 4.62~\cite{Ji2016} & 4.93~\cite{Ji2016} & \\
\hline
 & ($10\overline{1}1$) & 4.67 &  & 4.94~\cite{Ji2016} & 5.25~\cite{Ji2016} & \\
\hline
 & ($10\overline{1}2$) & 4.42 &  & 4.55~\cite{Ji2016} & 4.86~\cite{Ji2016} & \\
\hline
 & ($11\overline{2}1$) & 4.06 &  & 4.33~\cite{Ji2016} & 4.62~\cite{Ji2016} & \\
\hline
 & ($2\overline{1}\overline{1}2$) & 4.38 &  & 4.48~\cite{Ji2016} & 4.79~\cite{Ji2016} & \\
\hline
 & ($21\overline{3}0$) & 4.06 &  & 4.49~\cite{Ji2016} & 4.77~\cite{Ji2016} & \\
\hline
Ir & ($100$) & 5.54 & 5.42 & 5.55~\cite{Wang2014c} & 5.91~\cite{Waele2016} & 5.96~\cite{Derry2015}*\\
\hline
 & ($111$) & 5.42 & 5.18 & 5.5~\cite{Wang2014c} & 5.86~\cite{Waele2016} & 5.78~\cite{Derry2015}*\\
\hline
 & ($110$) & 5.01 & 4.83 & 4.96~\cite{Wang2014c} & 5.31~\cite{Waele2016} & 5.42~\cite{Derry2015}*\\
\hline
 & ($321$) & 5.0 &  & 5.07~\cite{Wang2014c} &  & 5.4~\cite{Kawano2008}*\\
\hline
 & ($211$) & 5.21 &  & 5.28~\cite{Wang2014c} &  & \\
\hline
 & ($310$) & 5.04 &  & 5.13~\cite{Wang2014c} &  & \\
\hline
 & ($210$) & 4.94 &  &  &  & 5.0~\cite{Kawano2008}*\\
\hline
 & ($331$) & 4.87 &  &  &  & 5.4~\cite{Kawano2008}*\\
\hline
Cd & ($0001$) & 3.63 & 3.76 & 3.81~\cite{Waele2016} & 4.21~\cite{Waele2016} & \\
\hline
 & ($10\overline{1}0$) & 3.72 &  & 4.08~\cite{Waele2016} & 4.49~\cite{Waele2016} & \\
\hline
 & ($21\overline{3}0$) & 3.76 &  & 3.9~\cite{Waele2016} & 4.21~\cite{Waele2016} & \\
\hline
Au & ($100$) & 5.0 & 4.9 & 5.07~\cite{Patra2017} & 5.49~\cite{Patra2017} & 5.22~\cite{Derry2015}*\\
\hline
 & ($111$) & 5.16 & 4.98 & 5.12~\cite{Patra2017} & 5.49~\cite{Patra2017} & 5.33~\cite{Derry2015}*\\
\hline
 & ($110$) & 4.93 & 4.88 & 4.94~\cite{Patra2017} & 5.36~\cite{Patra2017} & 5.16~\cite{Derry2015}*\\
\hline
 & ($321$) & 4.89 &  & 4.98~\cite{Wang2014c} &  & \\
\hline
 & ($211$) & 4.96 &  & 5.01~\cite{Wang2014c} &  & \\
\hline
 & ($310$) & 4.93 &  & 4.92~\cite{Wang2014c} &  & \\
\hline
 & ($210$) & 4.71 &  &  &  & 4.96~\cite{Kawano2008}*\\
\hline
 & ($311$) & 4.94 &  &  &  & 5.16~\cite{Kawano2008}*\\
\hline
Cu & ($111$) & 4.71 & 4.74 & 4.88~\cite{Patra2017} & 5.2~\cite{Patra2017} & 4.9~\cite{Derry2015}*\\
\hline
 & ($110$) & 4.2 & 4.32 & 4.38~\cite{Patra2017} & 4.68~\cite{Patra2017} & 4.56~\cite{Derry2015}*\\
\hline
 & ($100$) & 4.47 &  & 4.42~\cite{Patra2017} & 4.79~\cite{Patra2017} & 4.73~\cite{Derry2015}*\\
\hline
 & ($321$) & 4.22 &  & 4.35~\cite{Wang2014c} &  & \\
\hline
 & ($211$) & 4.22 &  & 4.45~\cite{Wang2014c} &  & 4.53~\cite{Kawano2008}*\\
\hline
 & ($310$) & 4.17 &  & 4.26~\cite{Wang2014c} &  & \\
\hline
 & ($210$) & 4.18 &  &  &  & 4.37~\cite{Kawano2008}*\\
\hline
 & ($311$) & 4.28 &  &  &  & 4.42~\cite{Kawano2008}*\\
\hline
Cr & ($110$) & 4.7 & 4.74 & 4.83~\cite{Waele2016} & 5.13~\cite{Waele2016} & \\
\hline
 & ($111$) & 4.04 &  & 4.09~\cite{Waele2016} & 4.39~\cite{Waele2016} & \\
\hline
 & ($100$) & 4.08 &  & 4.02~\cite{Waele2016} & 4.36~\cite{Waele2016} & \\
\hline
Zr & ($0001$) & 4.16 & 3.98 & 4.18~\cite{Waele2016} & 4.46~\cite{Waele2016} & \\
\hline
 & ($10\overline{1}0$) & 3.52 &  & 3.84~\cite{Waele2016} & 4.16~\cite{Waele2016} & \\
\hline
 & ($21\overline{3}0$) & 3.11 &  & 3.09~\cite{Waele2016} & 3.39~\cite{Waele2016} & \\
\hline
Li & ($100$) & 2.96 &  & 2.99~\cite{Wang2014c} & 3.13~\cite{Waele2016} & \\
\hline
 & ($110$) & 3.18 &  & 3.22~\cite{Wang2014c} & 3.36~\cite{Waele2016} & \\
\hline
 & ($111$) & 2.6 &  & 2.75~\cite{Wang2014c} & 2.92~\cite{Waele2016} & \\
\hline
 & ($210$) & 2.99 &  & 2.97~\cite{Wang2014c} &  & \\
\hline
 & ($331$) & 2.94 &  & 3.01~\cite{Wang2014c} &  & \\
\hline
 & ($311$) & 2.91 &  & 3.0~\cite{Wang2014c} &  & \\
\hline
Na & ($100$) & 2.53 &  & 2.64~\cite{Wang2014c} & 2.8~\cite{Waele2016} & \\
\hline
 & ($110$) & 2.71 &  & 2.84~\cite{Wang2014c} & 2.96~\cite{Waele2016} & \\
\hline
 & ($111$) & 2.5 &  & 2.58~\cite{Wang2014c} & 2.76~\cite{Waele2016} & \\
\hline
 & ($210$) & 2.65 &  & 2.68~\cite{Wang2014c} &  & \\
\hline
 & ($331$) & 2.49 &  & 2.73~\cite{Wang2014c} &  & \\
\hline
 & ($311$) & 2.45 &  & 2.69~\cite{Wang2014c} &  & \\
\hline
K & ($100$) & 2.15 &  & 2.22~\cite{Wang2014c} & 2.37~\cite{Waele2016} & \\
\hline
 & ($110$) & 2.37 &  & 2.37~\cite{Wang2014c} & 2.52~\cite{Waele2016} & \\
\hline
 & ($111$) & 2.13 &  & 2.18~\cite{Wang2014c} & 2.36~\cite{Waele2016} & \\
\hline
 & ($210$) & 1.95 &  & 2.23~\cite{Wang2014c} &  & \\
\hline
 & ($331$) & 2.29 &  & 2.29~\cite{Wang2014c} &  & \\
\hline
 & ($311$) & 2.15 &  & 2.24~\cite{Wang2014c} &  & \\
\hline
Rb & ($100$) & 2.1 &  & 2.12~\cite{Wang2014c} & 2.29~\cite{Waele2016} & \\
\hline
 & ($110$) & 2.23 &  & 2.24~\cite{Wang2014c} & 2.41~\cite{Waele2016} & \\
\hline
 & ($111$) & 2.05 &  & 2.1~\cite{Wang2014c} & 2.29~\cite{Waele2016} & \\
\hline
 & ($210$) & 1.94 &  & 2.13~\cite{Wang2014c} &  & \\
\hline
 & ($331$) & 2.18 &  & 2.18~\cite{Wang2014c} &  & \\
\hline
 & ($311$) & 2.08 &  & 2.13~\cite{Wang2014c} &  & \\
\hline
Cs & ($100$) & 1.97 &  & 1.97~\cite{Wang2014c} & 2.16~\cite{Waele2016} & \\
\hline
 & ($110$) & 2.04 &  & 2.07~\cite{Wang2014c} & 2.25~\cite{Waele2016} & \\
\hline
 & ($111$) & 1.59 &  & 1.97~\cite{Wang2014c} & 2.17~\cite{Waele2016} & \\
\hline
 & ($210$) & 1.77 &  & 2.0~\cite{Wang2014c} &  & \\
\hline
 & ($311$) & 1.65 &  & 2.0~\cite{Wang2014c} &  & \\
\hline
Ca & ($100$) & 2.56 &  & 2.76~\cite{Wang2014c} & 2.85~\cite{Waele2016} & \\
\hline
 & ($110$) & 2.74 &  & 2.81~\cite{Wang2014c} & 2.95~\cite{Waele2016} & \\
\hline
 & ($111$) & 2.94 &  & 2.94~\cite{Wang2014c} & 3.01~\cite{Waele2016} & \\
\hline
 & ($321$) & 2.52 &  & 2.78~\cite{Wang2014c} &  & \\
\hline
 & ($211$) & 2.55 &  & 2.83~\cite{Wang2014c} &  & \\
\hline
 & ($310$) & 2.39 &  & 2.71~\cite{Wang2014c} &  & \\
\hline
Sr & ($100$) & 2.47 &  & 2.47~\cite{Wang2014c} & 2.56~\cite{Waele2016} & \\
\hline
 & ($110$) & 2.49 &  & 2.54~\cite{Wang2014c} & 2.66~\cite{Waele2016} & \\
\hline
 & ($111$) & 2.53 &  & 2.57~\cite{Wang2014c} & 2.67~\cite{Waele2016} & \\
\hline
 & ($321$) & 2.33 &  & 2.51~\cite{Wang2014c} &  & \\
\hline
 & ($211$) & 2.37 &  & 2.45~\cite{Wang2014c} &  & \\
\hline
 & ($310$) & 2.31 &  & 2.47~\cite{Wang2014c} &  & \\
\hline
Ba & ($100$) & 2.28 &  & 2.31~\cite{Wang2014c} & 2.54~\cite{Waele2016} & \\
\hline
 & ($110$) & 2.35 &  & 2.38~\cite{Wang2014c} & 2.58~\cite{Waele2016} & \\
\hline
 & ($111$) & 2.24 &  & 2.29~\cite{Wang2014c} & 2.58~\cite{Waele2016} & \\
\hline
 & ($210$) & 1.9 &  & 2.35~\cite{Wang2014c} &  & \\
\hline
 & ($331$) & 2.35 &  & 2.34~\cite{Wang2014c} &  & \\
\hline
 & ($311$) & 2.21 &  & 2.33~\cite{Wang2014c} &  & \\
\hline
Be & ($0001$) & 5.32 &  & 5.29~\cite{Ji2016} & 5.45~\cite{Ji2016} & \\
\hline
 & ($10\overline{1}0$) & 4.48 &  & 4.52~\cite{Ji2016} & 4.71~\cite{Ji2016} & \\
\hline
 & ($10\overline{1}1$) & 4.95 &  & 5.03~\cite{Ji2016} & 5.23~\cite{Ji2016} & \\
\hline
 & ($10\overline{1}2$) & 4.71 &  & 4.81~\cite{Ji2016} & 5.04~\cite{Ji2016} & \\
\hline
 & ($11\overline{2}1$) & 4.35 &  & 4.58~\cite{Ji2016} & 4.82~\cite{Ji2016} & \\
\hline
 & ($2\overline{1}\overline{1}2$) & 4.79 &  & 4.81~\cite{Ji2016} & 4.94~\cite{Ji2016} & \\
\hline
 & ($21\overline{3}0$) & 4.14 &  & 4.17~\cite{Ji2016} & 4.38~\cite{Ji2016} & \\
\hline
Mg & ($0001$) & 3.61 &  & 3.76~\cite{Ji2016} & 3.89~\cite{Ji2016} & \\
\hline
 & ($10\overline{1}0$) & 3.39 &  & 3.64~\cite{Ji2016} & 3.76~\cite{Ji2016} & \\
\hline
 & ($10\overline{1}1$) & 3.64 &  & 3.7~\cite{Ji2016} & 3.88~\cite{Ji2016} & \\
\hline
 & ($10\overline{1}2$) & 3.58 &  & 3.63~\cite{Ji2016} & 3.74~\cite{Ji2016} & \\
\hline
 & ($11\overline{2}1$) & 3.4 &  & 3.56~\cite{Ji2016} & 3.68~\cite{Ji2016} & \\
\hline
 & ($2\overline{1}\overline{1}2$) & 3.62 &  & 3.67~\cite{Ji2016} & 3.8~\cite{Ji2016} & \\
\hline
 & ($21\overline{3}0$) & 3.35 &  & 3.49~\cite{Ji2016} & 3.72~\cite{Ji2016} & \\
\hline
Al & ($111$) & 4.0 &  & 4.2~\cite{Patra2017} & 4.97~\cite{Patra2017} & 4.32~\cite{Derry2015}*\\
\hline
 & ($100$) & 4.18 &  & 4.27~\cite{Patra2017} & 4.41~\cite{Patra2017} & 4.31~\cite{Derry2015}*\\
\hline
 & ($110$) & 4.0 &  & 3.96~\cite{Patra2017} & 4.08~\cite{Patra2017} & 4.23~\cite{Derry2015}*\\
\hline
Ga & ($001$) & 3.9 &  & 4.01~\cite{Waele2016} & 4.33~\cite{Waele2016} & \\
\hline
 & ($100$) & 4.19 &  & 4.22~\cite{Waele2016} & 4.56~\cite{Waele2016} & \\
\hline
 & ($010$) & 4.35 &  & 4.61~\cite{Waele2016} & 4.96~\cite{Waele2016} & \\
\hline
In & ($001$) & 3.74 &  & 3.88~\cite{Waele2016} & 4.22~\cite{Waele2016} & \\
\hline
 & ($100$) & 3.81 &  & 3.79~\cite{Waele2016} & 4.12~\cite{Waele2016} & \\
\hline
 & ($110$) & 3.81 &  & 3.92~\cite{Waele2016} & 4.3~\cite{Waele2016} & \\
\hline
Tl & ($0001$) & 3.5 &  & 3.56~\cite{Waele2016} & 3.98~\cite{Waele2016} & \\
\hline
 & ($10\overline{1}0$) & 3.42 &  & 3.56~\cite{Waele2016} & 3.98~\cite{Waele2016} & \\
\hline
 & ($21\overline{3}0$) & 3.34 &  & 3.52~\cite{Waele2016} & 3.94~\cite{Waele2016} & \\
\hline
 & ($21\overline{3}1$) & 3.27 &  & 3.52~\cite{Waele2016} & 3.95~\cite{Waele2016} & \\
\hline
C($P6_3/mmc$) & ($0001$) & 5.02 &  & 4.57~\cite{Waele2016} & 4.76~\cite{Waele2016} & 4.7~\cite{Kawano2008}*\\
\hline
C($Fd\overline{3}m$) & ($111$) & 4.61 &  & 4.36~\cite{Waele2016} & 4.54~\cite{Waele2016} & \\
\hline
 & ($100$) & 5.17 &  & 5.63~\cite{Waele2016} & 5.63~\cite{Waele2016} & \\
\hline
 & ($110$) & 5.22 &  & 5.3~\cite{Waele2016} & 5.56~\cite{Waele2016} & \\
\hline
Si & ($111$) & 4.64 &  & 4.67~\cite{Waele2016} & 4.81~\cite{Waele2016} & 4.6~\cite{Kawano2008}*\\
\hline
 & ($100$) & 4.79 &  & 4.56~\cite{Waele2016} & 4.71~\cite{Waele2016} & 4.91~\cite{Kawano2008}*\\
\hline
 & ($110$) & 4.97 &  & 5.0~\cite{Waele2016} & 5.23~\cite{Waele2016} & 4.8~\cite{Kawano2008}*\\
\hline
Ge & ($111$) & 4.52 &  & 4.38~\cite{Waele2016} & 4.55~\cite{Waele2016} & \\
\hline
 & ($100$) & 4.41 &  & 4.48~\cite{Waele2016} & 4.72~\cite{Waele2016} & \\
\hline
 & ($110$) & 4.51 &  & 4.74~\cite{Waele2016} & 4.99~\cite{Waele2016} & \\
\hline
Sn & ($111$) & 4.21 &  & 4.15~\cite{Waele2016} & 4.37~\cite{Waele2016} & \\
\hline
 & ($100$) & 4.38 &  & 4.23~\cite{Waele2016} & 4.49~\cite{Waele2016} & \\
\hline
 & ($110$) & 4.5 &  & 4.4~\cite{Waele2016} & 4.71~\cite{Waele2016} & \\
\hline
Pb & ($111$) & 3.68 &  & 3.76~\cite{Waele2016} & 4.08~\cite{Waele2016} & \\
\hline
 & ($100$) & 3.71 &  & 3.8~\cite{Waele2016} & 4.1~\cite{Waele2016} & \\
\hline
 & ($110$) & 3.6 &  & 3.73~\cite{Waele2016} & 4.03~\cite{Waele2016} & \\
\hline
As & ($0001$) & 4.44 &  & 4.41~\cite{Waele2016} & 4.71~\cite{Waele2016} & \\
\hline
Sb & ($0001$) & 4.23 &  & 4.24~\cite{Waele2016} & 4.5~\cite{Waele2016} & \\
\hline
 & ($10\overline{1}0$) & 4.29 &  & 4.39~\cite{Waele2016} & 4.72~\cite{Waele2016} & \\
\hline
 & ($11\overline{2}0$) & 4.24 &  & 4.41~\cite{Waele2016} & 4.62~\cite{Waele2016} & \\
\hline
Bi & ($0001$) & 3.98 &  & 4.07~\cite{Waele2016} & 4.37~\cite{Waele2016} & \\
\hline
 & ($11\overline{2}0$) & 3.98 &  & 4.14~\cite{Waele2016} & 4.39~\cite{Waele2016} & \\
\hline
Ti & ($0001$) & 4.38 &  & 4.42~\cite{Waele2016} & 4.67~\cite{Waele2016} & \\
\hline
 & ($10\overline{1}0$) & 3.51 &  & 3.9~\cite{Waele2016} & 4.23~\cite{Waele2016} & \\
\hline
 & ($21\overline{3}0$) & 3.21 &  & 3.19~\cite{Waele2016} & 3.45~\cite{Waele2016} & \\
\hline
 & ($21\overline{3}1$) & 3.26 &  & 3.51~\cite{Waele2016} & 3.8~\cite{Waele2016} & \\
\hline

\end{longtable}
\end{center}

\begin{minipage}{1\textwidth}
{\footnotesize
* See reference herein
}
\end{minipage}

\clearpage

\clearpage

\begin{minipage}{1\textwidth}
{\footnotesize
* See reference herein
}
\end{minipage}

\begin{center}
\begin{longtable}{ccccccc}
\caption{The R values for $\gamma_{\mbox{\scriptsize{hkl}}}$ and $\Phi_{\mbox{\scriptsize{hkl}}}$ as a function of normalized broken bonds per surface area ($\overline{\mbox{BB}}$) and $\Phi_{\mbox{\scriptsize{hkl}}}$ as a function of $\gamma_{\mbox{\scriptsize{hkl}}}$.} \label{fig:SchmoluchoskiCorrelation}\\

\hline Material &  spacegroup & \multicolumn{5}{c}{R values} \\
\toprule
\hline  &   & \multicolumn{2}{c}{$\gamma_{\mbox{\scriptsize{hkl}}}$ vs $\overline{\mbox{BB}}$}  & \multicolumn{2}{c}{$\Phi_{\mbox{\scriptsize{hkl}}}$ vs $\overline{\mbox{BB}}$} & \multicolumn{1}{c}{$\Phi_{\mbox{\scriptsize{hkl}}}$ vs $\gamma_{\mbox{\scriptsize{hkl}}}$} \\
\toprule
& & $1st NN$ & $2nd NN$ & $1st NN$ & $2nd NN$ & \\
\toprule
\hline
Cu & $Fm-3m$ & 0.99 & -- & -0.93 & -- & -0.93 \\
\hline
Er & $P6_3/mmc$ & -0.06 & 0.84 & 0.24 & -0.90 & -0.77 \\
\hline
La & $P6_3/mmc$ & 0.53 & 0.98 & -0.72 & -0.89 & -0.73 \\
\hline
Re & $P6_3/mmc$ & 0.57 & 0.79 & 0.02 & -0.89 & -0.28 \\
\hline
Ag & $Fm-3m$ & 0.96 & -- & -0.89 & -- & -0.88 \\
\hline
Ni & $Fm-3m$ & 1.00 & -- & -0.89 & -- & -0.85 \\
\hline
Ru & $P6_3/mmc$ & -0.40 & 0.95 & 0.64 & -0.88 & -0.91 \\
\hline
Ho & $P6_3/mmc$ & 0.35 & 0.75 & -0.24 & -0.88 & -0.70 \\
\hline
Zr & $P6_3/mmc$ & 0.52 & 0.73 & -0.15 & -0.87 & -0.48 \\
\hline
Nb & $Im-3m$ & 0.47 & 0.88 & -0.86 & -0.43 & -0.72 \\
\hline
Au & $Fm-3m$ & 0.92 & -- & -0.86 & -- & -0.74 \\
\hline
Eu & $P6_3/mmc$ & -0.02 & 0.77 & 0.20 & -0.84 & -0.74 \\
\hline
Tc & $P6_3/mmc$ & 0.80 & 0.92 & -0.18 & -0.84 & -0.42 \\
\hline
Y & $P6_3/mmc$ & -0.07 & 0.78 & 0.23 & -0.83 & -0.61 \\
\hline
Lu & $P6_3/mmc$ & -0.01 & 0.70 & 0.05 & -0.82 & -0.70 \\
\hline
Rh & $Fm-3m$ & 0.97 & -- & -0.82 & -- & -0.68 \\
\hline
W & $Im-3m$ & 0.94 & -0.00 & -0.81 & -0.52 & -0.65 \\
\hline
Tm & $P6_3/mmc$ & 0.34 & 0.69 & -0.15 & -0.80 & -0.47 \\
\hline
Li & $Im-3m$ & -0.38 & 0.77 & -0.34 & -0.79 & -0.43 \\
\hline
Sc & $P6_3/mmc$ & 0.67 & 0.67 & -0.23 & -0.79 & -0.66 \\
\hline
Ta & $Im-3m$ & 0.19 & 0.96 & -0.78 & -0.51 & -0.60 \\
\hline
Mo & $Im-3m$ & 0.47 & 0.21 & -0.78 & -0.53 & -0.55 \\
\hline
Pd & $Fm-3m$ & 0.95 & -- & -0.78 & -- & -0.71 \\
\hline
Cr & $Im-3m$ & 0.87 & 0.17 & -0.61 & -0.74 & -0.38 \\
\hline
Ti & $P6_3/mmc$ & 0.67 & 0.21 & 0.34 & -0.74 & 0.42 \\
\hline
V & $Im-3m$ & -0.27 & 0.89 & -0.74 & -0.67 & -0.40 \\
\hline
Cs & $Im-3m$ & 0.87 & 0.89 & -0.73 & -0.65 & -0.83 \\
\hline
Be & $P6_3/mmc$ & 0.45 & 0.88 & 0.43 & -0.73 & -0.15 \\
\hline
Hf & $P6_3/mmc$ & 0.80 & 0.82 & 0.11 & -0.73 & 0.07 \\
\hline
Ir & $Fm-3m$ & 0.93 & -- & -0.72 & -- & -0.50 \\
\hline
Co & $P6_3/mmc$ & 0.57 & 0.95 & -0.68 & -0.71 & -0.47 \\
\hline
Pt & $Fm-3m$ & 0.88 & -- & -0.71 & -- & -0.51 \\
\hline
Ca & $Fm-3m$ & 0.86 & -- & -0.70 & -- & -0.41 \\
\hline
Nd & $P6_3/mmc$ & 0.07 & 0.94 & 0.22 & -0.70 & -0.23 \\
\hline
Pr & $P6_3/mmc$ & 0.03 & 0.95 & 0.28 & -0.65 & -0.15 \\
\hline
Tl & $P6_3/mmc$ & 0.79 & 0.83 & -0.19 & -0.65 & -0.57 \\
\hline
Mg & $P6_3/mmc$ & 0.30 & 0.82 & -0.64 & -0.26 & -0.23 \\
\hline
Na & $Im-3m$ & 0.38 & 0.97 & -0.49 & -0.62 & -0.62 \\
\hline
Os & $P6_3/mmc$ & -0.10 & 0.94 & 0.24 & -0.60 & -0.30 \\
\hline
Sm & $P6_3/mmc$ & 0.10 & 0.70 & -0.32 & -0.58 & 0.09 \\
\hline
Sr & $Fm-3m$ & 0.85 & -- & -0.56 & -- & -0.46 \\
\hline
K & $Im-3m$ & 0.22 & 0.95 & -0.55 & -0.44 & -0.38 \\
\hline
Zn & $P6_3/mmc$ & 0.59 & 0.48 & -0.51 & 0.44 & -0.20 \\
\hline
Ce & $Fm-3m$ & 0.96 & -- & -0.49 & -- & -0.49 \\
\hline
Fe & $Im-3m$ & 0.40 & 0.23 & -0.44 & -0.48 & -0.49 \\
\hline
Pm & $P6_3/mmc$ & 0.46 & 0.88 & -0.40 & -0.44 & -0.21 \\
\hline
Yb & $Fm-3m$ & 0.76 & -- & -0.44 & -- & -0.23 \\
\hline
Rb & $Im-3m$ & 0.22 & 0.94 & -0.35 & -0.39 & -0.45 \\
\hline
Ba & $Im-3m$ & -0.01 & 0.98 & -0.20 & -0.04 & 0.01 \\
\hline
Pb & $Fm-3m$ & 0.94 & -- & -0.05 & -- & 0.02 \\
\hline
Cd & $P6_3/mmc$ & 0.53 & 0.51 & 0.01 & 0.41 & 0.24 \\
\hline
Al & $Fm-3m$ & 0.97 & -- & 0.22 & -- & 0.09 \\
\hline

\end{longtable}
\end{center}

\begin{minipage}{1\textwidth}
{\footnotesize
* See reference herein
}
\end{minipage}

\section*{Declarations of interest:}
None

\clearpage

\bibliography{workfunction_bibliography}

\providecommand{\latin}[1]{#1}
\providecommand*\mcitethebibliography{\thebibliography}
\csname @ifundefined\endcsname{endmcitethebibliography}
  {\let\endmcitethebibliography\endthebibliography}{}
\begin{mcitethebibliography}{49}
\providecommand*\natexlab[1]{#1}
\providecommand*\mciteSetBstSublistMode[1]{}
\providecommand*\mciteSetBstMaxWidthForm[2]{}
\providecommand*\mciteBstWouldAddEndPuncttrue
  {\def\EndOfBibitem{\unskip.}}
\providecommand*\mciteBstWouldAddEndPunctfalse
  {\let\EndOfBibitem\relax}
\providecommand*\mciteSetBstMidEndSepPunct[3]{}
\providecommand*\mciteSetBstSublistLabelBeginEnd[3]{}
\providecommand*\EndOfBibitem{}
\mciteSetBstSublistMode{f}
\mciteSetBstMaxWidthForm{subitem}{(\alph{mcitesubitemcount})}
\mciteSetBstSublistLabelBeginEnd
  {\mcitemaxwidthsubitemform\space}
  {\relax}
  {\relax}

\bibitem[Heimel \latin{et~al.}(2006)Heimel, Romaner, Bredas, and
  Zojer]{Heimel2006}
Heimel,~G.; Romaner,~L.; Bredas,~J.~L.; Zojer,~E. \emph{Physical Review
  Letters} \textbf{2006}, \emph{96}, 2--5\relax
\mciteBstWouldAddEndPuncttrue
\mciteSetBstMidEndSepPunct{\mcitedefaultmidpunct}
{\mcitedefaultendpunct}{\mcitedefaultseppunct}\relax
\EndOfBibitem
\bibitem[Lu \latin{et~al.}(2013)Lu, Hua, and Li]{Lu2013}
Lu,~H.; Hua,~G.; Li,~D. \emph{Applied Physics Letters} \textbf{2013},
  \emph{103}, 2619021--2619024\relax
\mciteBstWouldAddEndPuncttrue
\mciteSetBstMidEndSepPunct{\mcitedefaultmidpunct}
{\mcitedefaultendpunct}{\mcitedefaultseppunct}\relax
\EndOfBibitem
\bibitem[Lu \latin{et~al.}(2016)Lu, Liu, Yan, Li, Parent, and Tian]{Lu2016}
Lu,~H.; Liu,~Z.; Yan,~X.; Li,~D.; Parent,~L.; Tian,~H. \emph{Nature Publishing
  Group} \textbf{2016}, 1--11\relax
\mciteBstWouldAddEndPuncttrue
\mciteSetBstMidEndSepPunct{\mcitedefaultmidpunct}
{\mcitedefaultendpunct}{\mcitedefaultseppunct}\relax
\EndOfBibitem
\bibitem[Hua and Li(2016)Hua, and Li]{Hua2016}
Hua,~G.; Li,~D. \emph{Phys. Chem. Chem. Phys.} \textbf{2016}, \emph{18},
  4753--4759\relax
\mciteBstWouldAddEndPuncttrue
\mciteSetBstMidEndSepPunct{\mcitedefaultmidpunct}
{\mcitedefaultendpunct}{\mcitedefaultseppunct}\relax
\EndOfBibitem
\bibitem[Kawano(2008)]{Kawano2008}
Kawano,~H. \emph{Progress in Surface Science} \textbf{2008}, \emph{83},
  1--165\relax
\mciteBstWouldAddEndPuncttrue
\mciteSetBstMidEndSepPunct{\mcitedefaultmidpunct}
{\mcitedefaultendpunct}{\mcitedefaultseppunct}\relax
\EndOfBibitem
\bibitem[Li \latin{et~al.}(2015)Li, Rickman, and Schroeder]{Li2015a}
Li,~T.; Rickman,~B.~L.; Schroeder,~W.~A. \emph{Physical Review Special Topics -
  Accelerators and Beams} \textbf{2015}, \emph{18}, 1--11\relax
\mciteBstWouldAddEndPuncttrue
\mciteSetBstMidEndSepPunct{\mcitedefaultmidpunct}
{\mcitedefaultendpunct}{\mcitedefaultseppunct}\relax
\EndOfBibitem
\bibitem[Michaelson(1978)]{Michaelson1978}
Michaelson,~H.~B. \emph{IBM Journal of Research and Development} \textbf{1978},
  \emph{22}, 72--80\relax
\mciteBstWouldAddEndPuncttrue
\mciteSetBstMidEndSepPunct{\mcitedefaultmidpunct}
{\mcitedefaultendpunct}{\mcitedefaultseppunct}\relax
\EndOfBibitem
\bibitem[Miedema \latin{et~al.}(1973)Miedema, {De Boer}, and {De
  Chatel}]{Miedema1973}
Miedema,~A.~R.; {De Boer},~F.~R.; {De Chatel},~P.~F. \emph{Journal of Physics
  F: Metal Physics} \textbf{1973}, \emph{3}, 1558--1576\relax
\mciteBstWouldAddEndPuncttrue
\mciteSetBstMidEndSepPunct{\mcitedefaultmidpunct}
{\mcitedefaultendpunct}{\mcitedefaultseppunct}\relax
\EndOfBibitem
\bibitem[Smoluchowski(1941)]{Smoluchowski1941}
Smoluchowski,~R. \emph{Physical Review} \textbf{1941}, \emph{60},
  661--674\relax
\mciteBstWouldAddEndPuncttrue
\mciteSetBstMidEndSepPunct{\mcitedefaultmidpunct}
{\mcitedefaultendpunct}{\mcitedefaultseppunct}\relax
\EndOfBibitem
\bibitem[Sokolov and Jona(1984)Sokolov, and Jona]{Sokolov1984}
Sokolov,~J.; Jona,~F. \emph{Solid State Communications} \textbf{1984},
  \emph{49}, 307--312\relax
\mciteBstWouldAddEndPuncttrue
\mciteSetBstMidEndSepPunct{\mcitedefaultmidpunct}
{\mcitedefaultendpunct}{\mcitedefaultseppunct}\relax
\EndOfBibitem
\bibitem[Wang and Wang(2014)Wang, and Wang]{Wang2014c}
Wang,~J.; Wang,~S.~Q. \emph{Surface Science} \textbf{2014}, \emph{630},
  216--224\relax
\mciteBstWouldAddEndPuncttrue
\mciteSetBstMidEndSepPunct{\mcitedefaultmidpunct}
{\mcitedefaultendpunct}{\mcitedefaultseppunct}\relax
\EndOfBibitem
\bibitem[Ji \latin{et~al.}(2016)Ji, Zhu, and Wang]{Ji2016}
Ji,~D.~P.; Zhu,~Q.; Wang,~S.~Q. \emph{Surface Science} \textbf{2016},
  \emph{651}, 137--146\relax
\mciteBstWouldAddEndPuncttrue
\mciteSetBstMidEndSepPunct{\mcitedefaultmidpunct}
{\mcitedefaultendpunct}{\mcitedefaultseppunct}\relax
\EndOfBibitem
\bibitem[Brodie(1995)]{Brodie1995}
Brodie,~I. \emph{Physical Review B} \textbf{1995}, \emph{51},
  13660--13668\relax
\mciteBstWouldAddEndPuncttrue
\mciteSetBstMidEndSepPunct{\mcitedefaultmidpunct}
{\mcitedefaultendpunct}{\mcitedefaultseppunct}\relax
\EndOfBibitem
\bibitem[Wojciechowski and Borna(1997)Wojciechowski, and
  Borna]{Wojciechowski1997}
Wojciechowski,~K.~F.; Borna,~M. \emph{Vacuum} \textbf{1997}, \emph{48},
  257--259\relax
\mciteBstWouldAddEndPuncttrue
\mciteSetBstMidEndSepPunct{\mcitedefaultmidpunct}
{\mcitedefaultendpunct}{\mcitedefaultseppunct}\relax
\EndOfBibitem
\bibitem[Fazylov(2014)]{Fazylov2014}
Fazylov,~F. \emph{Philosophical Magazine} \textbf{2014}, \emph{94},
  1956--1966\relax
\mciteBstWouldAddEndPuncttrue
\mciteSetBstMidEndSepPunct{\mcitedefaultmidpunct}
{\mcitedefaultendpunct}{\mcitedefaultseppunct}\relax
\EndOfBibitem
\bibitem[Michaelson(1977)]{Michaelson1977}
Michaelson,~H.~B. \emph{Journal of Applied Physics} \textbf{1977}, \emph{48},
  4729--4733\relax
\mciteBstWouldAddEndPuncttrue
\mciteSetBstMidEndSepPunct{\mcitedefaultmidpunct}
{\mcitedefaultendpunct}{\mcitedefaultseppunct}\relax
\EndOfBibitem
\bibitem[Derry \latin{et~al.}(2015)Derry, Kern, and Worth]{Derry2015}
Derry,~G.~N.; Kern,~M.~E.; Worth,~E.~H. \emph{Journal of Vacuum Science {\&}
  Technology A: Vacuum, Surfaces, and Films} \textbf{2015}, \emph{33},
  060801\relax
\mciteBstWouldAddEndPuncttrue
\mciteSetBstMidEndSepPunct{\mcitedefaultmidpunct}
{\mcitedefaultendpunct}{\mcitedefaultseppunct}\relax
\EndOfBibitem
\bibitem[Helander \latin{et~al.}(2010)Helander, Greiner, Wang, and
  Lu]{Helander2010}
Helander,~M.~G.; Greiner,~M.~T.; Wang,~Z.~B.; Lu,~Z.~H. \emph{Applied Sur}
  \textbf{2010}, \emph{256}, 2602--2605\relax
\mciteBstWouldAddEndPuncttrue
\mciteSetBstMidEndSepPunct{\mcitedefaultmidpunct}
{\mcitedefaultendpunct}{\mcitedefaultseppunct}\relax
\EndOfBibitem
\bibitem[Singh-Miller and Marzari(2009)Singh-Miller, and
  Marzari]{Singh-Miller2009}
Singh-Miller,~N.~E.; Marzari,~N. \emph{Physical Review B - Condensed Matter and
  Materials Physics} \textbf{2009}, \emph{80}\relax
\mciteBstWouldAddEndPuncttrue
\mciteSetBstMidEndSepPunct{\mcitedefaultmidpunct}
{\mcitedefaultendpunct}{\mcitedefaultseppunct}\relax
\EndOfBibitem
\bibitem[Methfessel \latin{et~al.}(1992)Methfessel, Hennig, and
  Scheffler]{Methfessel1992}
Methfessel,~M.; Hennig,~D.; Scheffler,~M. \emph{Physical Review B}
  \textbf{1992}, \emph{46}, 4816--4829\relax
\mciteBstWouldAddEndPuncttrue
\mciteSetBstMidEndSepPunct{\mcitedefaultmidpunct}
{\mcitedefaultendpunct}{\mcitedefaultseppunct}\relax
\EndOfBibitem
\bibitem[Waele \latin{et~al.}(2016)Waele, Lejaeghere, Sluydts, and
  Cottenier]{Waele2016}
Waele,~S.~D.; Lejaeghere,~K.; Sluydts,~M.; Cottenier,~S. \emph{Physical Review
  B} \textbf{2016}, \emph{94}, 1--13\relax
\mciteBstWouldAddEndPuncttrue
\mciteSetBstMidEndSepPunct{\mcitedefaultmidpunct}
{\mcitedefaultendpunct}{\mcitedefaultseppunct}\relax
\EndOfBibitem
\bibitem[Skriver and Rosengaard(1992)Skriver, and Rosengaard]{Skriver1992}
Skriver,~H.~L.; Rosengaard,~N.~M. \emph{Physical Review B} \textbf{1992},
  \emph{46}, 7157--7168\relax
\mciteBstWouldAddEndPuncttrue
\mciteSetBstMidEndSepPunct{\mcitedefaultmidpunct}
{\mcitedefaultendpunct}{\mcitedefaultseppunct}\relax
\EndOfBibitem
\bibitem[Patra \latin{et~al.}(2017)Patra, Bates, Sun, and Perdew]{Patra2017}
Patra,~A.; Bates,~J.; Sun,~J.; Perdew,~J.~P. \emph{Arxiv} \textbf{2017},
  1--17\relax
\mciteBstWouldAddEndPuncttrue
\mciteSetBstMidEndSepPunct{\mcitedefaultmidpunct}
{\mcitedefaultendpunct}{\mcitedefaultseppunct}\relax
\EndOfBibitem
\bibitem[Durakiewicz \latin{et~al.}(2001)Durakiewicz, Arko, Joyce, Moore, and
  Halas]{Durakiewicz2001}
Durakiewicz,~T.; Arko,~A.; Joyce,~J.~J.; Moore,~D.~P.; Halas,~S. \emph{Physical
  Review B - Condensed Matter and Materials Physics} \textbf{2001}, \emph{64},
  1--8\relax
\mciteBstWouldAddEndPuncttrue
\mciteSetBstMidEndSepPunct{\mcitedefaultmidpunct}
{\mcitedefaultendpunct}{\mcitedefaultseppunct}\relax
\EndOfBibitem
\bibitem[Ald{\'{e}}n \latin{et~al.}(1995)Ald{\'{e}}n, Johansson, and
  Skriver]{Alden1995}
Ald{\'{e}}n,~M.; Johansson,~B.; Skriver,~H.~L. \emph{Physical Review B}
  \textbf{1995}, \emph{51}, 5386--5396\relax
\mciteBstWouldAddEndPuncttrue
\mciteSetBstMidEndSepPunct{\mcitedefaultmidpunct}
{\mcitedefaultendpunct}{\mcitedefaultseppunct}\relax
\EndOfBibitem
\bibitem[Stekolnikov \latin{et~al.}(2002)Stekolnikov, Furthmuller, Bechstedt,
  Furthm{\"{u}}ller, and Bechstedt]{Stekolnikov2002b}
Stekolnikov,~a.~A.; Furthmuller,~J.; Bechstedt,~F.; Furthm{\"{u}}ller,~J.;
  Bechstedt,~F. \emph{Physical Review B} \textbf{2002}, \emph{65}, 115318\relax
\mciteBstWouldAddEndPuncttrue
\mciteSetBstMidEndSepPunct{\mcitedefaultmidpunct}
{\mcitedefaultendpunct}{\mcitedefaultseppunct}\relax
\EndOfBibitem
\bibitem[Delchar(1986)]{Woodruff2016}
Delchar,~T.~A. \emph{{Modern Techniques of Surface Science}}, 3rd ed.;
  Cambridge University Press: Cambridge, 1986; p 464\relax
\mciteBstWouldAddEndPuncttrue
\mciteSetBstMidEndSepPunct{\mcitedefaultmidpunct}
{\mcitedefaultendpunct}{\mcitedefaultseppunct}\relax
\EndOfBibitem
\bibitem[Tran \latin{et~al.}(2016)Tran, Xu, Radhakrishnan, Winston, Sun,
  Persson, and Ong]{Tran2016a}
Tran,~R.; Xu,~Z.; Radhakrishnan,~B.; Winston,~D.; Sun,~W.; Persson,~K.~A.;
  Ong,~S.~P. \emph{Scientific Data} \textbf{2016}, \emph{3}, 1--13\relax
\mciteBstWouldAddEndPuncttrue
\mciteSetBstMidEndSepPunct{\mcitedefaultmidpunct}
{\mcitedefaultendpunct}{\mcitedefaultseppunct}\relax
\EndOfBibitem
\bibitem[Frank and Kasper(1958)Frank, and Kasper]{Frank1958}
Frank,~F.~C.; Kasper,~J.~S. \emph{Acta Crystallographica} \textbf{1958},
  \emph{11}, 184--190\relax
\mciteBstWouldAddEndPuncttrue
\mciteSetBstMidEndSepPunct{\mcitedefaultmidpunct}
{\mcitedefaultendpunct}{\mcitedefaultseppunct}\relax
\EndOfBibitem
\bibitem[Sun and Ceder(2013)Sun, and Ceder]{Sun2013a}
Sun,~W.; Ceder,~G. \emph{Surface Science} \textbf{2013}, \emph{617},
  53--59\relax
\mciteBstWouldAddEndPuncttrue
\mciteSetBstMidEndSepPunct{\mcitedefaultmidpunct}
{\mcitedefaultendpunct}{\mcitedefaultseppunct}\relax
\EndOfBibitem
\bibitem[Montoya and Persson(2017)Montoya, and Persson]{Montoya2017}
Montoya,~J.~H.; Persson,~K.~A. \emph{npj Computational Materials}
  \textbf{2017}, \emph{3}, 14\relax
\mciteBstWouldAddEndPuncttrue
\mciteSetBstMidEndSepPunct{\mcitedefaultmidpunct}
{\mcitedefaultendpunct}{\mcitedefaultseppunct}\relax
\EndOfBibitem
\bibitem[Ong \latin{et~al.}(2013)Ong, Richards, Jain, Hautier, Kocher, Cholia,
  Gunter, Chevrier, Persson, and Ceder]{Ong2013}
Ong,~S.~P.; Richards,~W.~D.; Jain,~A.; Hautier,~G.; Kocher,~M.; Cholia,~S.;
  Gunter,~D.; Chevrier,~V.~L.; Persson,~K.~A.; Ceder,~G. \emph{Computational
  Materials Science} \textbf{2013}, \emph{68}, 314--319\relax
\mciteBstWouldAddEndPuncttrue
\mciteSetBstMidEndSepPunct{\mcitedefaultmidpunct}
{\mcitedefaultendpunct}{\mcitedefaultseppunct}\relax
\EndOfBibitem
\bibitem[Jain \latin{et~al.}(2015)Jain, Ong, Chen, Medasani, Qu, Kocher,
  Brafman, Petretto, Rignanese, Hautier, Gunter, and Persson]{Jain2015b}
Jain,~A.; Ong,~S.~P.; Chen,~W.; Medasani,~B.; Qu,~X.; Kocher,~M.; Brafman,~M.;
  Petretto,~G.; Rignanese,~G.~M.; Hautier,~G.; Gunter,~D.; Persson,~K.~A.
  \emph{Concurrency Computation} \textbf{2015}, \emph{27}, 5037--5059\relax
\mciteBstWouldAddEndPuncttrue
\mciteSetBstMidEndSepPunct{\mcitedefaultmidpunct}
{\mcitedefaultendpunct}{\mcitedefaultseppunct}\relax
\EndOfBibitem
\bibitem[Mathew \latin{et~al.}(2017)Mathew, Montoya, Faghaninia, Dwarakanath,
  Aykol, Tang, Chu, Smidt, Bocklund, Horton, Dagdelen, Wood, Liu, Neaton, Ping,
  Persson, and Jain]{Mathew2017}
Mathew,~K. \latin{et~al.}  \emph{Computational Materials Science}
  \textbf{2017}, \emph{139}, 140--152\relax
\mciteBstWouldAddEndPuncttrue
\mciteSetBstMidEndSepPunct{\mcitedefaultmidpunct}
{\mcitedefaultendpunct}{\mcitedefaultseppunct}\relax
\EndOfBibitem
\bibitem[Winston \latin{et~al.}(2016)Winston, Montoya, and
  Persson]{Winston2016}
Winston,~D.; Montoya,~J.~H.; Persson,~K.~A. \emph{11th Gateway Computing
  Environments Conference} \textbf{2016},
  doi:10.6084/m9.figshare.4491623.v2\relax
\mciteBstWouldAddEndPuncttrue
\mciteSetBstMidEndSepPunct{\mcitedefaultmidpunct}
{\mcitedefaultendpunct}{\mcitedefaultseppunct}\relax
\EndOfBibitem
\bibitem[Cry(2016)]{CrystaliumWebsite}
Crystalium. \url{http://crystalium.materialsvirtuallab.org/}, 2016\relax
\mciteBstWouldAddEndPuncttrue
\mciteSetBstMidEndSepPunct{\mcitedefaultmidpunct}
{\mcitedefaultendpunct}{\mcitedefaultseppunct}\relax
\EndOfBibitem
\bibitem[MPW(2015)]{MPWebsite}
The Materials Project. \url{https://materialsproject.org/}, 2015\relax
\mciteBstWouldAddEndPuncttrue
\mciteSetBstMidEndSepPunct{\mcitedefaultmidpunct}
{\mcitedefaultendpunct}{\mcitedefaultseppunct}\relax
\EndOfBibitem
\bibitem[Haynes \latin{et~al.}(2017)Haynes, Lide, and Bruno]{CRC}
Haynes,~W.~M.; Lide,~D.~R.; Bruno,~T.~J. \emph{{CRC handbook of chemistry and
  physics}}, 97th ed.; CRC Press: Boca Raton, Fl, 2017\relax
\mciteBstWouldAddEndPuncttrue
\mciteSetBstMidEndSepPunct{\mcitedefaultmidpunct}
{\mcitedefaultendpunct}{\mcitedefaultseppunct}\relax
\EndOfBibitem
\bibitem[Rozkhov and Ye.(1971)Rozkhov, and Ye.]{Rozkhov1971}
Rozkhov,~O. K.~K.; Ye.,~E. \emph{{Rare Earth Metals and Alloys}}; Nauka:
  Moscow, 1971\relax
\mciteBstWouldAddEndPuncttrue
\mciteSetBstMidEndSepPunct{\mcitedefaultmidpunct}
{\mcitedefaultendpunct}{\mcitedefaultseppunct}\relax
\EndOfBibitem
\bibitem[Eastmen and Mee(1973)Eastmen, and Mee]{Eastmen1973}
Eastmen,~R.~M.; Mee,~C. H.~B. \emph{Journal of Physics F: Metal Physics}
  \textbf{1973}, \emph{3}, 1738--1745\relax
\mciteBstWouldAddEndPuncttrue
\mciteSetBstMidEndSepPunct{\mcitedefaultmidpunct}
{\mcitedefaultendpunct}{\mcitedefaultseppunct}\relax
\EndOfBibitem
\bibitem[Grepstad \latin{et~al.}(1976)Grepstad, Gartland, and
  Slagsvold]{Grepstad1976}
Grepstad,~J.; Gartland,~P.; Slagsvold,~B. \emph{Surface Science} \textbf{1976},
  \emph{57}, 348--362\relax
\mciteBstWouldAddEndPuncttrue
\mciteSetBstMidEndSepPunct{\mcitedefaultmidpunct}
{\mcitedefaultendpunct}{\mcitedefaultseppunct}\relax
\EndOfBibitem
\bibitem[Michaelides and Scheffler(2010)Michaelides, and
  Scheffler]{Michaelides2010}
Michaelides,~A.; Scheffler,~M. \emph{{\ldots} of Surface and Interface Science}
  \textbf{2010}, 1--40\relax
\mciteBstWouldAddEndPuncttrue
\mciteSetBstMidEndSepPunct{\mcitedefaultmidpunct}
{\mcitedefaultendpunct}{\mcitedefaultseppunct}\relax
\EndOfBibitem
\bibitem[Rahemi and Li(2015)Rahemi, and Li]{Rahemi2015}
Rahemi,~R.; Li,~D. \emph{Scripta Materialia} \textbf{2015}, \emph{99},
  41--44\relax
\mciteBstWouldAddEndPuncttrue
\mciteSetBstMidEndSepPunct{\mcitedefaultmidpunct}
{\mcitedefaultendpunct}{\mcitedefaultseppunct}\relax
\EndOfBibitem
\bibitem[Kiejna \latin{et~al.}(1979)Kiejna, Wojciechowski, and
  Zebrowksi]{Kiejna1979}
Kiejna,~A.; Wojciechowski,~K.~F.; Zebrowksi,~J. \emph{Journal of Physics F:
  Metal Physics} \textbf{1979}, \emph{9}, 1361\relax
\mciteBstWouldAddEndPuncttrue
\mciteSetBstMidEndSepPunct{\mcitedefaultmidpunct}
{\mcitedefaultendpunct}{\mcitedefaultseppunct}\relax
\EndOfBibitem
\bibitem[Ho and Bohnen(1987)Ho, and Bohnen]{Ho1987a}
Ho,~K.~M.; Bohnen,~K.~P. \emph{Physical Review Letters} \textbf{1987},
  \emph{59}, 1833--1836\relax
\mciteBstWouldAddEndPuncttrue
\mciteSetBstMidEndSepPunct{\mcitedefaultmidpunct}
{\mcitedefaultendpunct}{\mcitedefaultseppunct}\relax
\EndOfBibitem
\bibitem[Gaston \latin{et~al.}(2010)Gaston, Andrae, Paulus, Wedig, and
  Jansen]{Gaston2010a}
Gaston,~N.; Andrae,~D.; Paulus,~B.; Wedig,~U.; Jansen,~M. \emph{Phys. Chem.
  Chem. Phys.} \textbf{2010}, \emph{12}, 681--687\relax
\mciteBstWouldAddEndPuncttrue
\mciteSetBstMidEndSepPunct{\mcitedefaultmidpunct}
{\mcitedefaultendpunct}{\mcitedefaultseppunct}\relax
\EndOfBibitem
\bibitem[Fall \latin{et~al.}(1998)Fall, Binggeli, and Baldereschi]{Fall1998}
Fall,~C.~J.; Binggeli,~N.; Baldereschi,~A. \emph{Physical Review B}
  \textbf{1998}, \emph{58}, R7544--R7547\relax
\mciteBstWouldAddEndPuncttrue
\mciteSetBstMidEndSepPunct{\mcitedefaultmidpunct}
{\mcitedefaultendpunct}{\mcitedefaultseppunct}\relax
\EndOfBibitem
\bibitem[Lang and Kohn(1971)Lang, and Kohn]{Jolla1971}
Lang,~N.~D.; Kohn,~W. \emph{Physical Review B} \textbf{1971}, \emph{3},
  1215--1223\relax
\mciteBstWouldAddEndPuncttrue
\mciteSetBstMidEndSepPunct{\mcitedefaultmidpunct}
{\mcitedefaultendpunct}{\mcitedefaultseppunct}\relax
\EndOfBibitem
\end{mcitethebibliography}

\end{document}